\documentclass[12pt]{article}

\usepackage{amsmath}
\usepackage{amssymb}
\usepackage{amsfonts}
\usepackage{amsthm}
\usepackage{graphicx}
\newtheorem{remark}{Theorem}[section]
\newtheorem{propo}{Proposition}[section]

\usepackage{latexsym}
\usepackage{color,soul} 

\def\beq{\begin{eqnarray}}
\def\eeq{\end{eqnarray}}

\def\sign{\,\mbox{sgn}\,}

\renewcommand{\vec}[1]{{\bf #1}}

\def\al{\alpha}
\def\be{\beta}

\def\ga{\gamma}

\def\vp{\varepsilon}
\def\ep{\epsilon}

\def\ka{\kappa}
\def\la{\lambda}
\def\na{\nabla}
\def\pa{\partial}

\def\si{\sigma}

\def\th{\theta}

\def\Ga{\Gamma}

\def\Om{\Omega}

\begin{document}

\begin{center}

{\Large
Low-energy effects in a higher-derivative
gravity model
\\
with real and complex massive poles
}
\vskip 8mm

{\large Antonio Accioly$^{a}$,
\ \ \
Breno L. Giacchini$^{a}$,
\ \ {\large and} \ \
Ilya L. Shapiro$^{b,c}$}

\end{center}
\vskip 1mm

\begin{center}
{\sl
(a)
Centro Brasileiro de Pesquisas F\'{\i}sicas
\\
Rua Dr. Xavier Sigaud 150, Urca, 22290-180, Rio de Janeiro, RJ, Brazil
\vskip 3mm

(b) \ Departamento de F\'{\i}sica, ICE, Universidade Federal de Juiz de Fora,
\\
Campus Universit\'{a}rio,  Juiz de Fora, 36036-330, MG, Brazil
\vskip 3mm

(c) \ Tomsk State Pedagogical University and Tomsk State
University, Tomsk, Russia
}
\vskip 2mm\vskip 2mm

{\sl E-mails:
\ \
accioly@cbpf.br,
 \
breno@cbpf.br,
 \
shapiro@fisica.ufjf.br}

\end{center}
\vskip 6mm

\begin{quotation}
\noindent
\textbf{Abstract.} \
The most simple superrenormalizable model of quantum gravity
is based on the general local covariant six-derivative action. In
addition to graviton such a theory has massive scalar and tensor
modes. It was shown recently that in the case when the massive
poles emerge in complex conjugate pairs, the theory has also unitary
$S$-matrix and hence can be seen as a candidate to be a consistent
quantum gravity theory. In the present work we construct the modified
Newton potential and explore the gravitational light bending in a
general six-derivative theory, including the most interesting case of
complex massive poles. In the case of the light deflection the results
are obtained within classical and semiclassical approaches.
\vskip 3mm

\vskip 2mm

\vskip 2mm

\end{quotation}

\tableofcontents

\section{Introduction}
\label{S1}

Quantum gravity is an important part of a modern quantum field
theory (QFT) and of the gravitational physics. Since there are
relatively small chances to observe quantum corrections to the action
of gravity, one of the main targets of quantum gravity is to establish
the classical action capable of providing a consistent quantum
theory. Thinking in this direction we immediately realize
the relevant role played by higher-derivative terms, as well as
the difficult problem they represent. Even at the semiclassical level
one has to include  fourth derivative terms into the gravitational
action in order to provide renormalizability~\cite{UtDW} (see
also~\cite{birdav,book} for an introduction and~\cite{PoImpo}
for a recent
review), and the same is also true for the quantum theory of the
gravitational field itself~\cite{Stelle77}. On the other hand the
fourth derivative terms lead to ghost (or tachyonic ghost)
degrees of freedom in the physical spectrum of the theory, which implies
that there will be instabilities in the classical solutions. The existence
of a physically real ghost particle is a theoretical disaster: such a
particle has negative kinetic energy, therefore it will accelerate
emitting plenty of gravitons. As a result the absolute value of its
negative energy rapidly goes to infinity and an infinitely powerful
gravitational explosion will occur. Since nothing of this sort was
observed so far, the problem should have some theoretical resolution.

During many years the discussions about the problem of ghosts
were based on the following approaches:

{\it i)} \  Treating all higher derivatives,
together with the corresponding quantum contributions, as
small corrections~\cite{Simon-90}, in the same way as it is done
in QED to avoid the run-away solutions~\cite{LL-4}. Within this
approach one has to ignore a great difference which exists between
gravity and QED, since the latter is renormalizable without higher
derivative terms. As a consequence, in QED the higher derivative
terms are not running, and one can always assume that they are just
part of a more general action which is presumably non-local and free
of ghosts. A very simple example of an artificial ghost appearance
was recently discussed in the context of effective approach to
quantum field theory \cite{Burgess}. In the case of gravity the
assumption of smallness of the higher-derivative terms is much
more {\it ad hoc} and is certainly suitable only at the energy scales
much below the Planck scale. A natural question is why we need a
theory of quantum gravity which works only at low energies, and
this question remains without answer.

Another alternative is to assume that the higher derivative ghosts
exist only as virtual  excitations, but for some unknown reason
they are not generated as physical particles at the sub-Planck
energies. The creation of a Planck-mass ghost from vacuum
requires a concentration of gravitons with Planck density, and in
some gravity models this may be impossible~\cite{DG}. A strong
support for this hypothesis comes from the low-energy stability of
the classical cosmological solutions in higher derivative gravity
models \cite{GW-Stab,HD-Stab,GW-HD-MPLA}.

{\it ii)} \  In string theory the space-time metric is regarded as an
effective composite field and one can redefine it in such a way
that the ghost degrees of freedom disappear~\cite{zwei}. Formally
this solves the problem\footnote{It is worth noting that the same
can be achieved in the semiclassical theory of gravity.}.
However, there are two difficulties in this approach.
\ {\it First}, the
procedure is ambiguous. For instance one can remove or not
$R^2$, $R^3$, and other similar terms, or one can just modify
their coefficients. All of these terms do not contribute to ghosts,
but at the same time they do affect classical gravitational
solutions~\cite{maroto}, giving rise to a great uncertainty in the
predictions of the theory.
\ {\it Second}, the removal of all the terms which produce ghosts,
such as $R_{\mu\nu}\Box^k R^{\mu\nu}$ for $k \geq 0$ and
$R\Box^k R$  for $k \geq 1$ must be performed with absolute
precision. Any infinitesimal deviation from zero in any of these
coefficients means that the ghost comes back and that its mass
is huge, even compared to the Planck mass. Then the effect of such
a ghost (e.g., instability of Minkowski space) will be even stronger
and not weaker, as one can imagine.

Furthermore, at lower energies our experience shows that the
appropriate description of quantum effects is within the QFT,
not string theory. However, the loop corrections within QFT
typically break down an absolutely precise fine-tuning which is
required to avoid ghosts (see, e.g., the discussion in
\cite{CountGhost}). Therefore, string theory is helpful in solving
ghost problems only if we assume that all the low-energy quantum
physics, in all its details, is a consequence of a string theory, that
provides the requested cancellations. Such an assumption looks
very strong and certainly difficult to believe in without further arguments.

{\it iii)} \  In the framework of four derivative quantum gravity
one can assume that the dressed gravitational
propagator, with quantum corrections, makes the ghost unstable.
Then the theory could possibly have a unitary $S$-matrix. This idea
was nicely introduced in~\cite{Tomb-77,salstr,Antomb}, but the final
conclusion was that the information which can be obtained via the
perturbative QFT approaches is not sufficient to decide whether this
mechanism is working or not~\cite{Johnston}.

{\it iv)} \ Another possibility is to start from a non-local theory with
infinitely many derivatives of the metric. One can consider
non-local form factors such as
$\,R_{\mu\nu}\Phi(\Box) R^{\mu\nu}\,$ and $\,R\Psi(\Box) R\,$
 so that no other poles will exist in the propagator besides the
 massless one corresponding to the graviton. This procedure can
 be applied either in string theory~\cite{Tseytlin-95} as an
 alternative to the metric reparametrization of~\cite{zwei}, or
 in quantum gravity~\cite{Tomboulis-97}
 (see also~\cite{modesto,Modesto2016,SRQG-beta} for recent
 developments and further references). The main disadvantage of
 this approach is
 that the functions $\,\Phi(\Box)\,$ and $\,\Psi(\Box)\,$ must
 be chosen with absolute precision. As a result the ghost-free
conditions can not survive any kind of low-energy quantum
corrections~\cite{CountGhost}. After the specially tuned form
of the form factors gets modified, there is an infinite amount of
ghost-like states, all of them corresponding to complex poles.

{\it v)} \ The last possibility is to consider local gravitational theories
with more than four derivatives. These theories have remarkable
quantum properties. Typically they are superrenormalizable
\cite{highderi} and also, in case of massive complex poles, can be
unitary in the Lee-Wick sense \cite{LM-Sh}. Therefore, these theories
are capable to solve the conflict between UV renormalizability and
unitarity in quantum gravity. Regardless of remaining problems,
these models are unitary without any sort of fine-tuning and
hence they represent simpler alternatives to the non-local models.

Of course, at the present level the higher derivative theories with
complex massive poles can not be seen as a complete solution of
the quantum gravity problem, but they look as strong candidates.
Therefore, it makes sense to explore their IR properties at the
classical level and identify observables which might be useful for
experimental detection of higher derivatives.
The model
of our interest is the simplest theory which admits complex poles,
with the action of the form
\beq
\label{Lag6orderGravity}
\mathcal{S} = \int d^4 x \sqrt{-g}
\Big\lbrace &&
\dfrac{2}{\kappa^2} R
+ \dfrac{\alpha}{2} R^2 + \dfrac{\beta}{2} R_{\mu\nu}^2
+ \dfrac{A}{2} R \Box R
+ \dfrac{B}{2} R_{\mu\nu} \square R^{\mu\nu}
+ \mathcal{L}_M \Big\rbrace\,,
\eeq
where $\mathcal{L}_M$ is the matter Lagrangian,
$\ka^2/2 = 16 \pi G = M_P^{-2}$, $G$ is Newton's constant and
$M_P$ is the reduced Planck mass. $\al$, $\be$, $A$ and $B$ are
free parameters, the first two being dimensionless, and $A$ and $B$
carry dimension of (mass)$^{-2}$. The values of these parameters
should be determined by experimental data.

Let us note that the structure of the poles in the dressed propagator 
of gravitons was considered in some recent publications, for example
in~\cite{Ref1,Ref2,Ref3}, where physical effects of complex poles were 
discussed. In particular, one of the results of \cite{Ref1} is that the 
perturbative unitarity can be restored by the resummation. In general, 
the approach of the present work differs from the one in these 
references since we regard the higher-derivative model as a
fundamental rather than effective. In the IR sector considered
here, however, the difference between the approaches is supposed 
to be irrelevant, as heavy degrees of freedom should decouple 
in the long-distance limit. In our present work this is not 
the case nonetheless, because we are partially dealing with the propagation of 
massive degrees of freedom up to the cosmic scales, or at least up 
to the scale of laboratory. Indeed, since there are no direct 
experiments on quantum gravity, very different approaches to the 
problem should be seen as legitimate in this area.

The model~\eqref{Lag6orderGravity} is the particular case of the
superrenormalizable quantum gravity theory formulated in
\cite{highderi}. One can generalize it by adding
${\cal O}(R^3_{\dots})$-terms to the
action, but these terms should be irrelevant for our purposes, since
we are interested in the effects related to the linear gravitational
perturbations.

In the present paper we will explore in detail the two most
obvious low-energy observables which can be used to falsify the
presence of fourth- and six-derivative terms in the theory
(\ref{Lag6orderGravity}). The first part of the work is about
the modified Newtonian potential. If compared to the previous
works on the subject
(see, e.g., \cite{Quandt-Schmidt,modNew,Giacchini-poles}),
we include here the cases of complex and multiple real poles,
that provides a better perspective and understanding for
the modified potential in general polynomial higher derivative
models.

The second part of the paper is devoted to the bending of light in
the theories with higher derivatives. This issue is attracting a
great deal of attention, especially in relation to quantum
gravity and quantum field theory effects. Indeed, quantum effects
can be partially taken into account in the low-energy domain by the
use of semiclassical methods. In the case of gravity let us mention,
e.g., the influence of the one-loop vacuum polarization in the
propagation of photons on a curved background. This issue was
explored in the papers \cite{Berends&Gastmans} and
\cite{Drummond&Hathrell} using two different approaches. In the
former work the effect is described by the differential cross section.
It gives the correct leading term for the gravitational bending angle
of an unpolarized beam plus a semiclassical correction, which depends
on the energy of the photons.
On the other hand, in~\cite{Drummond&Hathrell} the semiclassical
correction is introduced in the interaction potential between an
external gravitational field and a photon. As a result the deflection
angle depends on the photon's polarization, but it is non-dispersive.
According to~\cite{Drummond&Hathrell}, this version of the
semiclassical consideration is the correct one, since it assumes that
for macroscopic systems the photon is better described by a
compact wave packet with a definite path in the gravitational field.
In Sec.~\ref{S5} one can find the discussion of this issue in the
context of higher-derivative gravity. In particular, we elaborate
on the explanation concerning the limits of applicability of the
semiclassical approach similar to the one of \cite{Berends&Gastmans},
and explain why the method based on the cross sections usually can
not be used to describe the bending of light at astronomical scales.

The bending of light in the theory \eqref{Lag6orderGravity} is
briefly discussed in the parallel work~\cite{Seesaw} which is
devoted to the possibility of a specific seesaw mechanism in higher
derivative quantum gravity. The much more detailed treatment of
this issue here complements the discussion of the parallel work.

The paper is organized as follows. In Sec.~\ref{S2} a
generalization of a theorem by Teyssandier \cite{Teyssandier89}
for the six-order gravity is formulated. The theorem, which is
proved in Appendix~\ref{S7}, presents
the general solution for the linearized sixth-order gravity as a
linear combination of five auxiliary fields.
In Sec.~\ref{S3} we
study the
modified Newtonian potential of the theory. The poles of  the
propagator can be either real (simple or degenerate) or complex.
In particular, we show that the potential is regular at the
origin, extending the result of~\cite{modNew}\footnote{A more general treatment
of this issue is given in the parallel work~\cite{Giacchini-poles}.}.
Section~\ref{S4} is devoted to the study of the classical gravitational
deflection of light rays, for each of the possible types of poles.
The quantum mechanical formulation of the scattering process
and the restricted applicability of such an approach to macroscopic
systems is discussed in Sec.~\ref{S5}.
In Sec.~\ref{S6} we draw our conclusions.

Our notations are as follows. The units correspond to
$\hbar = c = 1$. The signature is
$\eta_{\mu\nu} = \text{diag} (1,-1,-1,-1)$, and the Riemann
and Ricci tensors are
\beq
{R^\rho}_{\la\mu\nu}
= \pa_\mu \Ga^\rho_{\la\nu}
- \pa_\nu \Ga^\rho_{\la\mu}
+ \Ga^\si_{\la\nu} \Ga^\rho_{\si\mu}
- \Ga^\si_{\la\mu} \Ga^\rho_{\si\nu}
\label{curvas}
\eeq
and $R_{\mu\nu} = R^\rho_{\,\,\mu\nu\rho}$.
This choice of notations is intended to facilitate the comparison
of our calculations with the previous work \cite{Accioly15}
on the four-derivative gravity.

\section{Field generated by a point-like mass in rest}
\label{S2}

It is clear that the right choice of fields parametrization and of a
suitable gauge condition may lead to an essential simplification of
the field equations. This is especially important for the higher-order
gravity models, which have rather complicated dynamical equations.
In 1989 Teyssandier~\cite{Teyssandier89} introduced a useful form
of the third-order coordinate condition for the linearized
fourth-order gravity described by the action
\beq
\label{Lag4orderGravity}
\mathcal{S}_4
&=&
\int d^4 x \sqrt{-g} \left\{ \dfrac{2}{\kappa^2} R
+ \dfrac{\alpha}{2} R^2 + \dfrac{\beta}{2} R_{\mu\nu}^2
+ \mathcal{L}_M \right\} \,,
\eeq

In the Teyssandier gauge the
general solution of the linearized field equations are written as
a linear combination of three decoupled fields~\cite{Teyssandier89}.
In terms of these
auxiliary fields the weak gravitational field generated by a static
source can be promptly computed, as well as the classical potential
of the theory, which is proportional to the $(00)$-component of
the metric perturbation.

Our goal is to obtain similar representation in the framework
of the sixth-order gravity model \eqref{Lag6orderGravity}. The
variational principle applied to the action $\mathcal{S}[g_{\mu\nu}]$
leads to the field equations of the sixth-order gravity:
\beq
\label{eqMotion6}
&& \dfrac{2}{\kappa^2} \left( R_{\mu\nu} - \frac{R}{2} g_{\mu\nu} \right)
+ \dfrac{\alpha}{2} \left[ 2RR_{\mu\nu} + 2 \nabla_{\mu} \nabla_{\nu} R
- 2g_{\mu\nu} \square R - \frac{R^2}{2} g_{\mu\nu} \right]
\nonumber
\\
&+&
\dfrac{\beta}{2} \left[ - \frac{1}{2} g_{\mu\nu} R_{\alpha\beta}^2
+ \nabla_{\mu} \nabla_{\nu} R
+ 2 R_{\mu\sigma\rho\nu}R^{\sigma\rho}
- \frac{1}{2} g_{\mu\nu} \square R
- \square R_{\mu\nu} \right]
\nonumber
\\
&+&
\dfrac{A}{2} \bigg[ R_{\mu\nu} \square R
+ R \square R_{\mu\nu}
+ 2 \square\nabla_{\mu} \nabla_{\nu} R
- 2g_{\mu\nu} \square^2 R - (\nabla_\mu R) (\nabla_\nu R)
+ \frac{1}{2} g_{\mu\nu} (\nabla R)^2 \bigg]
\nonumber
\\
&+&
\dfrac{B}{2} \bigg[ \square \nabla_{\mu} \nabla_{\nu} R
+ 2 \square (R_{\mu\sigma\rho\nu}R^{\sigma\rho})
 - \square^2 \big(R_{\mu\nu}  + \frac{1}{2} g_{\mu\nu} R\big)
- 4 (\nabla_\sigma R_{\nu\rho})(\nabla^\sigma {R_\mu}^\rho)
\nonumber
\\
&+&
2(\nabla_\sigma R_{\rho(\nu})(\nabla_{\mu)} R^{ \sigma\rho})
+ R_{\nu\sigma}\nabla_\rho\nabla_\mu R^{\rho\sigma}
+ R_{\mu\sigma}\nabla_\rho\nabla_\nu R^{\rho\sigma}
+ \frac{1}{2} g_{\mu\nu} (\nabla_\lambda R_{\rho\sigma})^2
\nonumber
\\
&-&
\left( \nabla^\sigma R + 2 R^{\rho\sigma} \nabla_\rho \right)
 \nabla_{(\mu} R_{\nu)\sigma}
  - (\nabla_\mu R_{\rho\sigma})(\nabla_\nu R^{\rho\sigma})
- 2 R_{\sigma(\mu} \square {R_{\nu)}}^{\sigma}
\bigg]
= \,-\, \frac{1}{2}\,T_{\mu\nu}\,,
\eeq
where the parenthesis in the indices denote symmetrization,
e.g.,
\beq
\nabla_{(\mu} R_{\nu)\sigma} \equiv \frac{1}{2}
\left( \nabla_{\mu} R_{\nu\sigma} + \nabla_{\nu} R_{\mu\sigma}\right)\,.
\nonumber
\eeq

In the weak field regime the metric can be considered as a
fluctuation around the flat space,
\beq
g_{\mu\nu} = \eta_{\mu\nu} + \kappa h _{\mu\nu}\,,
\eeq
with $\vert \kappa h _{\mu\nu} \vert \ll 1$. The Ricci tensor
$R_{\mu\nu}$ and the scalar curvature $R$ up to the first
order in $\ka$ are
\beq
\label{RicciLin}
R^{(1)}_{\mu\nu}
&=& \frac{\kappa}{2} \left[ \square h_{\mu\nu}
- \eta^{\lambda\rho} (\gamma_{\lambda\mu,\nu\rho}
+ \gamma_{\lambda\nu,\mu\rho}) \right] \,,
\eeq
\beq
\label{RLin}
R^{(1)}
&=& \kappa \left( \frac{1}{2} \square h - \eta^{\lambda\rho}
\eta^{\mu\nu} \gamma_{\lambda\mu,\nu\rho} \right) \,.
\eeq
In the last expressions we used the notations
\beq
\label{Linearizados}
\ga_{\mu\nu} \,=\, h_{\mu\nu}
- \frac{1}{2} \eta_{\mu\nu} h\,,
\qquad
h \,=\, \eta^{\mu\nu} h_{\mu\nu}.
\eeq

Since the equations of motion are already expanded to the order
$\,\ka^2$, the d'Alembertian is calculated using the flat metric,
$\,\square =\eta^{\mu\nu} \pa_\mu \pa_\nu$.

Using the expressions \eqref{RicciLin}-\eqref{Linearizados},  the
linearized  equations of motion~\eqref{eqMotion6} are

\beq
\label{eqMotion6LinA}
&&
\left( \frac{2}{\kappa^2} - \frac{\beta}{2} \square
- \frac{B}{2} \square^2\right) \left( R^{(1)}_{\mu\nu}
- \frac{1}{2} \eta_{\mu\nu} R^{(1)} \right)
- \left( \alpha + \frac{\beta}{2}
+ A \square + \frac{B}{2} \square \right)
\left( \eta_{\mu\nu} \square R^{(1)}
- \partial_\mu \partial_\nu R^{(1)} \right)
\nonumber
\\
&& \,=\, - \frac{1}{2}\,T_{\mu\nu}\,.
\eeq

The trace of Eq.~\eqref{eqMotion6LinA} has the form
\beq
\label{traceEq}
\left( \alpha + \frac{\beta}{2} + A \square
+ \frac{B}{2} \square \right) \square R^{(1)}
= - \frac{1}{3} \left( \frac{2}{\kappa^2}
- \frac{\beta}{2} \square
- \frac{B}{2} \square^2\right) R^{(1)}
+ \frac{1}{6}\,T \,.
\eeq
Replacing \eqref{traceEq} into \eqref{eqMotion6LinA} yields
\beq
&& \left( \frac{2}{\kappa^2} - \frac{\beta}{2} \square
- \frac{B}{2} \square^2\right) \left( R^{(1)}_{\mu\nu}
- \frac{1}{6} \eta_{\mu\nu} R^{(1)} \right)
+ \left( \alpha
+ \frac{\beta}{2} + A \square + \frac{B}{2} \square \right)
\partial_\mu \partial_\nu R^{(1)}
\nonumber
\\
&& \,=\,
\frac{1}{6}\,T \eta_{\mu\nu} - \frac{1}{2}\,T_{\mu\nu}\,.
\eeq
Inserting the expression \eqref{RicciLin} for the first order
Ricci tensor into the preceding equation we obtain
\beq
\left[\frac{\ka^2}{4}\,\big(\be +  B\square\big)\square - 1\right]
\left(\square h_{\mu\nu}
- \dfrac{1}{3\kappa}\,R^{(1)} \eta_{\mu\nu} \right)
\,+\,  \Ga_{(\mu,\nu)}
\,=\,
2\left( T_{\mu\nu} - \dfrac{1}{3}\,T \eta_{\mu\nu}\right) ,
\label{eqMotion6LinB}
\eeq
where we defined the quantities
\beq
\Ga_{\mu}
&=&
\left( 1 - \dfrac{\kappa^2\beta}{4}\square
- \dfrac{\kappa^2 B}{4}\square^2 \right)
{\ga_{\mu\rho}}^{,\rho}
 - \dfrac{\kappa}{2} \left( \alpha + \dfrac{\beta}{2}
+ A \square + \dfrac{B}{2} \square \right) R^{(1)}_{,\mu} \,.
\eeq

Hence, implementing  the gauge condition  $\,\Ga_\mu = 0\,$
makes the problem of solving the linearized field equations
 \eqref{eqMotion6LinA} for $h_{\mu\nu}$ to be equivalent to
 the system consisting of the gauge condition and of Eq.
 \eqref{eqMotion6LinB}.  The convenience of this gauge is
to allow the solution to be expressed in terms of auxiliary fields.
Let us formulate this statement as a Theorem, with the proof
postponed to Appendix~\ref{S7}.

\begin{remark} \label{Theorem1}
The general solution of the system constituted by \eqref{eqMotion6LinB} 
and the gauge condition $\Gamma_{\mu}=0$ 
can be presented in the form
\beq
\label{Coro_GeneralSolution}
h_{\mu\nu}
&=&
h_{\mu\nu}^{(E)} + \Psi_{\mu\nu} + \bar{\Psi}_{\mu\nu}
- \eta_{\mu\nu} \Phi - \eta_{\mu\nu} \bar{\Phi}\,,
\eeq
where the auxiliary fields $\,h_{\mu\nu}^{(E)}$,
$\Psi_{\mu\nu}$, $\bar{\Psi}_{\mu\nu}$,
$\Phi$ and $\bar{\Phi}$ satisfy the second order equations
\beq
\label{Coro_h(E)}
\square h_{\mu\nu}^{(E)}
&=&
\dfrac{\kappa}{2}
\Big( \dfrac{1}{2}\, T\eta_{\mu\nu} - T_{\mu\nu} \Big) \,,
\label{Coro_hE_1}
\\
{\gamma_{\mu\nu}^{(E),\nu}} = 0 , \qquad
&&
\gamma_{\mu\nu}^{(E)} \,\equiv \,
h_{\mu\nu}^{(E)} - \dfrac{1}{2} \eta_{\mu\nu} h^{(E)}\,,
\label{Coro_hE_2}
\\
\label{Coro_Psi}
( m_{2+}^2 + \square ) \Psi_{\mu\nu}
&=&
\dfrac{\kappa}{2} \Big( T_{\mu\nu}
- \dfrac{1}{3} T\eta_{\mu\nu} \Big)\, ,
\label{Coro_Psi_1}
\\
(m_{2-}^2 + \square ) \bar{\Psi}_{\mu\nu}
&=&
m_{2+}^2 \Psi_{\mu\nu} \,,
\label{Coro_Psi_2}
\\
\left(\Psi_{\mu\nu}
\,+\,
\bar{\Psi}_{\mu\nu}\right)^{,\mu\nu}
&=&
\square \left(\Psi + \bar{\Psi}\right)\,,
\label{Coro_Psi_3}
\\
\label{Coro_Phi}
( m_{0+}^2 + \square ) \Phi
&=&
\dfrac{\ka}{12}\,T\,,
\label{Coro_Psi1}
\\
( m_{0-}^2 + \square ) \bar{\Phi}
&=&
 m_{0+}^2 \Phi \,.
 \label{Coro_Psi2}
\eeq
\end{remark}

Here and in what follows  we use the condensed notations
\beq
\Psi = \eta^{\mu\nu} \Psi_{\mu\nu}\,,
\quad
\sigma_1 = 3\alpha + \beta\,, \quad
\sigma_2 = 3A+B
\label{sigmas}
\eeq
and
\beq
\label{Def_masses}
m_{2\pm}^2 \,=\, \dfrac{-\frac{\beta \vert B \vert}{B}
\pm \sqrt{\be^2 + \frac{16}{\ka^2}B}}{- 2 \vert B \vert}\,, \quad
m_{0\pm}^2 \,=\,
\dfrac{\frac{\si_1 \vert \si_2 \vert}{\si_2}
\pm \sqrt{\si_1^2 - \frac{8}{\ka^2}\si_2}}{2\vert\si_2\vert}\,.
\eeq

According to the Theorem~\ref{Theorem1} formulated above, it is
possible to split the field $h_{\mu\nu}$ into a linear combination
of the five fields: a massless tensor representing the solution of
linearized  Einstein's equations in de Donder gauge, two massive
tensor fields  $\Psi_{\mu\nu}$ and $\bar{\Psi}_{\mu\nu}$ and two
scalars $\Phi$ and $\bar{\Phi}$.  Let us stress that in the
present case the massive fields with the same spin are not dynamically
independent. For this reason, as it will be shown in the next section,
the theory under discussion has a finite modified Newtonian
potential, regardless of the (complex or real) nature of the
quantities $m_{2\pm}$ and $m_{0\pm}$.

Using the previous theorem it is straightforward to calculate
the field generated by a point-like mass in rest at $\textbf{r} = 0$.
The corresponding energy-momentum tensor is
$T_{\mu\nu}(\textbf{r}) = M \eta_{\mu 0} \eta_{\nu 0}
\,\delta^{(3)}(\textbf{r})$. The solution for
$\,h_{\mu\nu}^{(E)}\,$ is the same as in Einstein's
gravity in the de Donder gauge:
\beq
\label{solutionEinstein}
h_{\mu\nu}^{(E)}(\textbf{r})
= \frac{M\kappa}{16\pi r}\, \left( \eta_{\mu\nu}
- 2\eta_{\mu 0}\,\eta_{\nu 0} \right) \,.
\eeq

The solutions for the massive tensor fields read
\beq
\label{solutionPsi}
\Psi_{\mu\nu}(\textbf{r})
&=&
\frac{M \ka}{8 \pi}
\left( \eta_{\mu 0}\,\eta_{\nu 0}
- \frac{1}{3} \eta_{\mu\nu} \right)
\,\frac{ e^{-m_{2+}r}}{r}
\eeq
and
\beq
\label{solutionPsiBarra}
\bar{\Psi}_{\mu\nu}(\textbf{r})
&=&
\frac{M \kappa}{8 \pi}  \left( \eta_{\mu 0}\eta_{\nu 0}
- \frac{1}{3} \eta_{\mu\nu} \right) \frac{m_{2+}^2}{m_{2+}^2
- m_{2-}^2}
 \left( \frac{e^{-m_{2-}r}}{r}
- \frac{e^{-m_{2+}r}}{r} \right) \,.
\eeq
It is easy to verify that these solutions satisfy the subsidiary
gauge condition \eqref{Coro_Psi_3}.

For the scalar modes we have
\beq
\label{solutionPhi}
\Phi(\textbf{r})
&=&
\frac{M\kappa}{48\pi} \,\frac{e^{-m_{0+}r}}{r},
\\
\label{solutionPhiBarra}
\bar{\Phi}(\textbf{r})
&=&
\frac{M\kappa}{48\pi}\, \frac{m_{0+}^2}{m_{0+}^2 - m_{0-}^2}
\left( \frac{e^{-m_{0-}r}}{r} - \frac{e^{-m_{0+}r}}{r} \right)\,.
\eeq
By  inserting the last five expressions into Eq.
\eqref{Coro_GeneralSolution} one finds the non-zero components
of the metric, $\,h_{00}\,$ and $\,h_{11} = h_{22} = h_{33}$, in the
form
\beq
\label{h00}
h_{00}(\textbf{r})
&=&
 \frac{M\kappa}{16\pi} \left(  - \frac{1}{r} + \frac{4}{3} F_2
 - \frac{1}{3} F_0 \right),
\\
\label{h11}
h_{11}(\textbf{r})
&=&
\frac{M\kappa}{16\pi} \left(  - \frac{1}{r} + \frac{2}{3} F_2
+ \frac{1}{3}F_0 \right)\, ,
\eeq
where ($k = 0,2$ labels the spin of the particle)
\beq
F_k
=
\frac{m_{k+}^2}{m_{k+}^2 - m_{k-}^2} \frac{e^{-m_{k-}r}}{r}
 + \frac{m_{k-}^2}{m_{k-}^2 - m_{k+}^2} \frac{e^{-m_{k+}r}}{r}
\,.
\label{Fk}
\eeq

Equations~\eqref{h00} and~\eqref{h11} represent the weak field generated
by a point mass in the general sixth-order gravity. In the previous
work~\cite{modNew} the $(00)$-component of the metric
perturbation has been computed in the more general case
containing terms $\square^n$ of arbitrary order in the action, but
only for real and non-degenerate massive poles of the propagator.
The expressions~\eqref{h00} and~\eqref{h11} apply to all types of
poles. 


\section{Modified Newtonian potential in the sixth-order gravity}
\label{S3}

The modified Newtonian potential (we shall use simply ``potential'' in what
follows) of the sixth-order gravity
can be directly read off from the solution~\eqref{h00} for the
field generated by a point-like mass in rest,
\beq
V(r)
&=&
\frac{\ka}{2}\,h_{00}(r)
\,=\,
MG \left(- \frac{1}{r} + \frac{4}{3} F_2 - \frac{1}{3} F_0\right)\,,
\eeq
with the functions $F_{0,2}$ defined in Eq.~\eqref{Fk}.

In this section we analyse the possible types of ``masses'' allowed
by the sixth-order gravity and their influence on the potential.
The calculations require only $h_{00}$; the other components of the metric will prove relevant
in the further sections dedicated to the gravitational light deflection.
The relevant quantities to be analysed are $F_k$, hence the results
will also be useful later on.

The following observation is in order. Complex massive poles
are not allowed in the fourth-order gravity, since they would imply
non-physical complex values for the potential.
However, in the sixth-order gravity the massive
modes of the same spin form dynamically dependent pairs. As we
shall see in short, this makes complex poles admissible and leads
to a real potential with oscillatory modes.



\subsection{Real poles}
\label{ss31}

In what follows we explore three different possibilities for real
poles, namely, pairs of different poles, including the special
situation in which one of the poles is much heavier than the
other, and the case of multiple (degenerate) poles.

\subsubsection{Real simple poles}
\label{sss311}

Real simple poles occur in the propagator of the massive tensor
field provided that
\beq
\label{real,condition2}
\beta < 0 ,
\quad
B < 0,
\quad
\beta^2  + \frac{16B}{\kappa^2} > 0\,,
\eeq
which enables one to redefine $m_{2\pm}^2$ as
\beq
m_{2\pm,\text{real}}^2
&=&
\dfrac{\beta \pm \sqrt{\be^2 + \frac{16}{\ka^2}B}}{2 B}\,.
\eeq
These masses satisfy the condition $m_{2-} > m_{2+}$, where
the lightest one corresponds to the well-known ghost mode and
the other is a healthy particle~\cite{highderi}.

With respect to the scalar field, the conditions $m_{0\pm}^2 > 0$
and $m_{0+} \neq m_{0-}$ yield
\beq
\label{real,condition0}
\sigma_1
=
3\al + \beta > 0 \,,
\quad
\si_2 = 3 A + B > 0,
\quad
(3\alpha + \beta)^2 - \frac{8(3 A + B)}{\kappa^2}> 0\,.
\eeq
Under these conditions one can redefine the scalar masses as
\beq
m_{0\pm,\text{real}}^2
&=&
 \dfrac{\si_1 \pm \sqrt{\si_1^2
- \frac{8\sigma_2}{\kappa^2}}}{2 \sigma_2}.
\eeq
Note that if \eqref{real,condition2} holds, then $\al$ and
$A$ must be positive. For the scalar field $m_{0+} > m_{0-}$,
but now the largest mass corresponds to the ghost
mode~\cite{highderi,modNew}. The reason for the qualitative
difference between the scalar and tensor cases is that in the latter
there exists the graviton, which is a healthy massless particle.

The expression for the potential is
\beq
V_\text{real}(r)
&=&
-\, \frac{ MG}{r} \,+\, \frac{4\, MG}{3}
 \left(  \frac{m_{2+}^2}{m_{2+}^2
 - m_{2-}^2} \frac{e^{-m_{2-}r}}{r}
 + \frac{m_{2-}^2}{m_{2-}^2
 - m_{2+}^2} \frac{e^{-m_{2+}r}}{r} \right)
\nonumber
\\
&& - \frac{ MG}{3} \left( \frac{m_{0+}^2}{m_{0+}^2
- m_{0-}^2} \frac{e^{-m_{0-}r}}{r}
+ \frac{m_{0-}^2}{m_{0-}^2
- m_{0+}^2} \frac{e^{-m_{0+}r}}{r} \right) ,
\eeq
which is just a particular case of the result obtained in
Ref.~\cite{modNew} by means of a different technique.
This potential is regular at the origin. Our following
calculations will show that this feature is also present if
the massive poles are degenerate or complex.

\subsubsection{Real degenerate poles}
\label{3.1.2}

The condition for having degenerate poles in the propagator of
the tensor or scalar fields is, respectively,
\beq
B = - \frac{\beta^2\kappa^2}{16}
\quad
\mbox{and}
\quad
\sigma_2  = \frac{\sigma_1^2\kappa^2}{8}\,.
\eeq
These formulas correspond to transforming the last inequalities
in~Eqs.~\eqref{real,condition2} and
\eqref{real,condition0} into equalities. Thus, the masses $m_{k}$
are defined by $\,m_2 = \sqrt{\beta/(2B)}$ and $m_0
= \sqrt{\sigma_1 / (2\sigma_2)}$.

It proves useful to consider this situation starting from the assumption
that the difference between the two real masses is small,
\beq
m_{2-} &=& m_{2+} + \epsilon_2 \,=\, m_2 + \epsilon_2
\nonumber
\\
m_{0+} &=& m_{0-} + \epsilon_0 \,=\, m_0 + \epsilon_0
\,,
\label{masses close}
\eeq
with $0< \epsilon_k / m_k \ll 1$.
Then the quantity $F_k$ reads
\beq
F_k
&=&
\left( - \frac{m_k}{2\epsilon_k} + \frac{1}{4}
 - \frac{\epsilon_k}{8m_k} \right) \frac{e^{-(m_k+\epsilon_k)r}}{r}
+
\left( \frac{m_k}{2\epsilon_k}
+ \frac{3}{4} + \frac{\epsilon_k}{8m_k} \right)
\frac{e^{-m_k r}}{r}
+ \mathcal{O}\left( \frac{\epsilon_k^2}{m_k^2}\right) .
\label{Fk_degenerate_eps}
\eeq

The limit $\epsilon_k \to 0$ is smooth, and
we arrive at the expression for $F_k$ for real degenerate poles,
\beq
\label{Fk_degenerate}
F_k
&\longrightarrow &
\left( \frac{1}{r} + \frac{m_k}{2} \right) e^{-m_k r}\,.
\eeq
The potential for two pairs of degenerate real poles assumes the form
\beq
\label{eq42}
V_\text{degen}(r)
&=&
MG \bigg[  - \dfrac{1}{r}
+ \dfrac{4}{3} \left( \dfrac{1}{r} + \dfrac{m_2}{2} \right)
e^{-m_2 r}\,
-
\dfrac{1}{3} \left( \dfrac{1}{r} + \dfrac{m_0}{2} \right) e^{-m_0 r}
\bigg] \,,
\eeq
which is indeed finite at the origin,
\beq
V_\text{degen}(0) &=&
- \,\frac{MG}{3} \left( 2 m_2 - \frac{m_0}{2} \right) \,.
\eeq
The result~\eqref{eq42} is in agreement with~\cite{Quandt-Schmidt}, where it was
considered the particular case $\, \beta = B = 0$.

\subsubsection{Real poles with strong hierarchy}
\label{3.1.3}

Another possibility allowed by the sixth-order gravity is to have one
of the masses of the auxiliary fields some (or many) orders of
magnitude smaller than the other:
\beq
m_{2-} 
\gg m_{2+}
\quad
\mbox{and/or}
 \quad
 m_{0+} 
 \gg m_{0-}.
\label{hiera}
\eeq
This situation leads to
potentially observable effects of higher derivatives at low
energies, e.g., through modifications of inverse-square force
low which could be detected in laboratory
experiments\footnote{Another
consequence of this possibility is related to the alleged
protection against Ostrogradsky-type instabilities \cite{HD-Stab,GW-HD-MPLA}, which would be less
efficient.}.
The possibility of such a strong hierarchy is discussed
in detail in the parallel paper~\cite{Seesaw}, which is
mainly devoted to this issue in general higher-derivative
gravities. Hence we will give here just a brief comment.
The  conditions (\ref{hiera}) can be achieved, respectively,
provided that $16 |B| \ll \kappa^2 \beta^2$ and/or that
$8 \sigma_2 \ll \kappa^2 \sigma_1^2$. It is easy to see
that if both conditions hold, in the leading order in $m_{2+}/m_{2-}$ (and $m_{0-}/m_{0+}$)
the potential reduces to the approximate form
\beq
\label{PotentialSpecialCase}
V_4(r)
&=&
MG \left(  - \frac{1}{r} + \frac{4}{3} \frac{e^{-m_{2+}r}}{r}
- \frac{1}{3}  \frac{e^{-m_{0-}r}}{r} \right)  .
\eeq
As it should be expected, this expression coincides with that obtained in
Ref.~\cite{Stelle77} within the fourth-order gravity, i.e.,
the theory defined by the action~\eqref{Lag4orderGravity}.
Qualitatively, this means that at longer distances the heaviest
masses have no effect.

Let us remember that the only possibility of reducing the lightest
masses in~\eqref{hiera} is to increase the coefficients $\al$ and $\be$ of the
fourth-derivative terms. In other words, tuning the
sixth-order coefficients do not reduce the lightest masses.
Further results on the viability of a gravitational seesaw-like
mechanism can be found in~\cite{Seesaw}.

\subsection{Complex poles}
\label{ss32}

Complex poles in the propagator of the spin-2 field can occur
provided that
\beq
\beta^2 + \frac{16B}{\ka^2} < 0 \,,
\quad
\left(\be^2 + \frac{16B}{\ka^2} \right)^{1/2}\, = \, i c_2\,,
\label{c2}
\eeq
with $c_2>0$ for definiteness. The first condition requires
$\,B<0$, while the four-derivative parameter $\,\be\,$ can
be  either positive or negative---differently from the real poles case,
Eq.~\eqref{real,condition2}.

The positions of the poles are defined by
\beq
m_{2\pm}^2
&=&
\frac{\beta \pm ic_2}{2B}\,.
\eeq
The square root of these quantities yield the ``masses''
\beq
\label{complexmassM2}
m_{2+}
&=&
a_2 - i b_2
\quad
\mbox{and}
\quad
m_{2-} \,=\, a_2 + i b_2 \,,
\eeq
where $a_2,b_2 > 0$ are defined through
\beq
\label{a2_and_c2_def}
a_2^2
&=&
\frac{-\beta + \sqrt{\beta^2 + c_2^2}}{4 \vert B \vert}
\,=\,
\frac{-\be + \sqrt{\frac{16\vert B \vert}{\ka^2}}}{4 \vert B \vert }\,,
\nonumber
\\
b_2^2
&=&
\frac{\beta + \sqrt{\beta^2 + c_2^2}}{4 \vert B \vert}
\,=\,
\frac{\be + \sqrt{\frac{16\vert B \vert}{\ka^2}}}{4 \vert B \vert }\,.
\eeq

One can always assume that $m_{2+}$ and $m_{2-}$ have positive
real parts. A short comment is in order here. If choosing $m_{2\pm}$
with negative real part in Eq.~\eqref{complexmassM2}, then the
decreasing real exponentials would turn to be increasing, introducing
into the potential growing oscillating modes at large distances. To
avoid these growing modes one would have to choose growing
exponentials as solution of the system~\eqref{Coro_Psi_1}-\eqref{Coro_Psi_3}. In this case
the negative real part of the ``masses'' would combine with the
increasing exponentials yielding decreasing oscillatory modes,
resulting precisely in Eq.~\eqref{m2_complex_potential} below.
Hence, the generality is not lost due to our choice of signs.

Finally, replacing \eqref{complexmassM2} into the
expression for $F_2$ leads us to
\beq
\label{m2_complex_potential}
F_2
&=&
\left[ \cos (b_2 r)
- \frac{\beta}{c_2} \sin (b_2 r) \right]  \frac{e^{-a_2 r}}{r}\,,
\eeq
which is a real quantity.

The condition for complex ``masses'' in the scalar field reads
\beq
\sigma_1^2 - \frac{8\sigma_2}{\kappa^2} < 0
\quad
\Longrightarrow
\quad
\sigma_2 > 0, \quad \sigma_1 \in \mathbb{R}\,.
\eeq
Similar to the spin-2 case, we define
\beq
i c_0 \,=\, \sqrt{\sigma_1^2 - \frac{8\sigma_2^2}{\kappa^2}}
\,,\quad
m_{0\pm} = a_0 \pm i b_0 \,,
\eeq
where ($a_0,b_0 > 0$)
\beq
a_0^2 = \frac{\sigma_1
+ \sqrt{\frac{8\sigma_2}{\kappa^2}}}{4 \sigma_2},
\quad
b_0^2
= \frac{-\sigma_1
+ \sqrt{\frac{8\sigma_2}{\kappa^2}}}{4 \sigma_2}
\, .
\eeq
The contribution of the scalar field to the potential is
\beq
\label{m0_complex_potential}
F_0
&=&
\left[\cos (b_0 r) + \frac{\sigma_1}{c_0}
\sin (b_0 r) \right]  \frac{e^{-a_0 r}}{r}\,.
\eeq

Taking together the contributions~\eqref{m2_complex_potential}
and~\eqref{m0_complex_potential} we arrive at the potential
in the case of complex poles,
\beq
\label{ComplexPotential}
V_\text{C}(r) &=&
-\, \frac{MG}{r}
+ \frac{4\,MG}{3} \left[ \cos (b_2 r)
- \frac{\beta}{c_2} \sin (b_2 r) \right]
\frac{e^{-a_2 r}}{r}
\nonumber
\\
&-&
\frac{MG}{3} \left[ \cos (b_0 r)
+ \frac{\sigma_1}{c_0} \sin (b_0 r) \right]
\frac{e^{-a_0 r}}{r} \,.
\eeq
It is straightforward to verify that this potential is finite at
$r=0$. Indeed,  this feature can be extended to the theory
of arbitrary order in the derivatives, including the case of
multiple complex poles~\cite{Giacchini-poles}.

The main distinguished feature of the complex poles case is the
presence of oscillating terms.
Depending on which quantity is greater in the pair $(a_k, b_k)$,
the oscillatory terms can be more or less relevant in the potential.
For example, in the case of the spin-2 field, $\beta < 0$ implies
$a_2 > b_2$. Since the characteristic length of the Yukawa
potential is $\,2\pi/a_2\,$ and the period of the oscillating terms
is $\,2\pi/b_2$, the oscillations can be smooth, yielding an
appreciable contribution only at distances larger than the Yukawa
length $\,2\pi/a_2$. There, the potential associated to this field
has an oscillating sign, but with a small absolute value due to the
suppression caused by $a_2$. Hence, at these distances the
potential is dominated by the Newtonian term owed to the
graviton.

On the other hand, if $\beta > 0$ it follows that $a_2 < b_2$.
Then the space period of oscillations is typically smaller than the
range of the Yukawa factor. This situation implies a significant
change in the behaviour of the potential at small distances, with the
contribution of the spin-2 field changing its sign. The same
argument applies \textit{mutatis mutandis} to the scalar field.
In case $\,a_k = 0\,$ the ``masses'' are purely imaginary
quantities which correspond to tachyonic modes. In this case
$F_k$ loses its damping term yielding a non-Newtonian
behaviour in the infinity. It is clear that this case can be ruled
out.

It is noteworthy that when we allowed massive complex poles, the
constraints on $\beta$ and $\sigma_1$ were relaxed. As we have
just mentioned, important changes in the ultraviolet behaviour of
the potential occur if, contrary to the real mass case, it is chosen
$\be \geq 0$ and/or $\si_1 \leq 0$. The case of $\be = 0$ and/or
$\sigma_1 = 0$ makes the real and imaginary parts $a_k$ and
$b_k$ to assume the same value, hence only the cosine functions
remain in the expression for the potential\footnote{When the
first version of the present work was under preparation, we
learned that the potential for the particular case $\be = \al = 0$
and $a_2 = a_0$ was derived in Ref.~\cite{Modesto2016}.}.

Likewise the case of real poles, one might suppose a sort of natural
seesaw mechanism which could reduce $a_k$ and $b_k$ and bring
the phenomenology of those modes to the low-energy scale. Indeed,
this can only happen for unnatural values of the massive parameters
at the action. Namely, in order to have $a_2 \approx b_2 \ll M_P$
one has to impose $\vert B \vert \gg M_P^{-2}$. A more detailed
and general discussion on this subject can be found in Ref.~\cite{Seesaw}.

Before closing this section, let us return to the Theorem~\ref{Theorem1}.
In Sec. \ref{S2} it was mentioned that the auxiliary fields of
the same spin have coupled dynamics. At the same time,
the equations for spin-2 and spin-0 components are factorized.
Due to this fact the cancelation of the Newtonian singularity occurs
independently of the (complex or real) nature and the multiplicity
of the massive poles. In brief, such a cancelation takes place if
there is at least one massive state in each of the sectors~\cite{Giacchini-poles}.

\section{Light bending: classical approach}
\label{S4}

Up to this point all the discussions were related to the
$(00)$-component of the metric~\eqref{h00}. Here and in the
following section we shall use this and also other  components
to study the weak-field regime of the gravitational deflection of
light within the sixth-order gravity. This issue has already been
analyzed in the framework of the fourth-derivative theory
(see~\cite{Accioly15} and references therein). In these works
the phenomenon of light bending was
used to derive a lower-bound on the mass of the tensor  mode
of the metric. The sixth-order model which we deal with here
has a richer variety of possible scenarios. The main purpose of
our present study is to systematically explore all of them for
different types of poles in the gravitational propagator.

The gravitational light bending problem has been explored
by using both classical or semiclassical approaches. In several
works it was explained that these two methods may lead to
different results
(see, e.g., \cite{Accioly15,Delbourgo&Phocas,Berends&Gastmans,
Caldwell,Drummond&Hathrell,Holstein}).
In the present section we analyse the phenomenon from a classical
point of view, that is, by treating both gravity and light as classical
fields. In the next section we describe the semiclassical
approach and discuss its applicability, so as to explain the
mentioned difference.

In order to arrive at a better understanding of the qualitative
features  of the gravitational deflection of a light ray passing
close to a massive body we shall use the so-called
$\ep$-$\mu$-form of Maxwell equations in curved
space-time~\cite{LightmanLee,FischbachFreeman}. This formalism
can be applied to static, spherically symmetric gravitational fields,
since under such circumstances it is always possible to find a
coordinate system where the metric has the isotropic form
\beq
\label{SSSMetric}
&&
g_{00} = g_{00}(r)
, \quad
g_{0i}=0
, \quad
g_{ij}=-\delta_{ij}f(r),
\eeq
for some function $f(r)$, where $r= |\vec{r}|$.
Using this metric it is not difficult to show  that the
inhomogeneous Maxwell equations
\beq
\label{Maxwell1}
{F^{\mu\nu}}_{;\mu} = J^\nu
\eeq
can be cast into the form
\beq
\label{Maxwell2}
\na \cdot (\vp \textbf{E} ) = \rho\,,
\quad
\frac{\pa}{\pa t} \big(\vp \textbf{E} \big)
- \na\times \left(  \frac{\textbf{B}}{\mu}  \right)  =
\textbf{j},
\eeq
where
\beq
\ep = \mu = \sqrt{\frac{f(r)}{g_{00}(r)}}.
\label{ep}
\eeq
These equations have the form of the usual (flat-space) Gauss's
and Amp\`{e}re's laws in a medium with refractive index
\beq
\label{index}
n(r)
&=&
\sqrt{\epsilon\mu} \,=\, \sqrt{\frac{f(r)}{g_{00}(r)}}\,\,.
\eeq

Thus, in the geometric optics limit, i.e., if the wavelength of light
is much smaller than the curvature scale, the influence of gravity on
light can be taken into account through \eqref{Maxwell2}, which can
be naturally interpreted as if gravity endows the flat space-time with
an effective refractive index. For example, the deflection of a light
ray passing close to a massive body can then be evaluated using the
Snell-Descartes law. Following the calculations of~\cite{FischbachFreeman},
the deflection angle $\theta$ for
a light ray passing in the vicinity of a massive body with the impact
parameter $\rho$ is given, to the first order in $G$, by the expression
\beq
\label{Angle}
\theta &=& -\,
\int_{-\infty}^{+\infty}  \frac{\rho}{rn(r)} \frac{dn(r)}{dr} dx\,,
\quad
\text{where} \quad
r = \sqrt{x^2 + \rho^2}
\eeq
and the trajectory of the photon is parametrized by $x$.
\
A small observation concerning the limits of integration
in~\eqref{Angle} is on order.
According to the scheme introduced in~\cite{FischbachFreeman}
the integration is performed starting at the position $x$ of the
light source (a distant star, for example), up to the position of the
observer, respectively to the massive scattering object. Since we
consider deflection caused by the Sun, it is natural to suppose that
both the light source and the Earth correspond to the space
infinities. However, for a precise calculation in more exotic
scenarios (e.g., those with complex poles), the upper limit related
to  Earth's position may need to be redefined.

Since the field generated by a point-like mass in rest found
in Sec.~\ref{S2} is already in the isotropic form, it is
straightforward to evaluate the effective refractive index
associated to the sixth-order gravity. From
Eqs.~\eqref{h00} and \eqref{h11} it follows that, to the
first order,
\beq
n(r)
& = &
\sqrt{\frac{1-\kappa h_{11}(r)}{1+\kappa h_{00}(r)}}
\nonumber
\\
& = &
1 - MG \left( - \frac{1}{r} + \frac{4}{3} F_2 - \frac{1}{3} F_0\right)
 - MG \left( - \frac{1}{r} + \frac{2}{3} F_2 + \frac{1}{3} F_0\right)
\nonumber
\\
& = & n_{\text{GR}}(r) - 2 MG F_2\,,
\label{Ref4}
\eeq
where
\beq
n_{\text{GR}}(r) \equiv 1 + \frac{2 M G}{r}
\eeq
is the effective refractive index of general relativity.

The immediate conclusion which follows from the expression
\eqref{Ref4} is that light bending in this theory does not depend
directly on the scalar excitations $m_{0\pm}$, and hence on the
sectors $R^2$ and $R\square R$. This result is rather expected,
since both sectors can be regarded as the result of a conformal
transformation on the weak-field metric. Since the curved-space
Maxwell equations are conformally invariant, these terms have
no direct effect on the light deflection, in the leading
approximation. Let us note that the semiclassical derivation of
the same statement for $R+R^2$ can be found in \cite{BD0}
and will be extended to the theory with $R\square R$ term
in the next section.

On the other hand, the scalar modes may have an indirect
influence on the bending of light, through the
redefinition of Newton's constant $G$ and the related
calibration of mass of astronomical bodies. This effect
is typical in the literature
on the masless Brans-Dicke theory
\cite{Will-book}\footnote{An important consideration
 concerning the effective Newton constant in metric-scalar
 models, including cosmological aspects of the problem, was
given in~\cite{PolStar2000}.}. The situation for the massive
Brans-Dicke theory can be very different, as explained, e.g.,
in the Refs.~\cite{massiveBD,Will-2012}.

Indeed,  it is even easier to understand the difference between
massless and massive cases for the model of $R+\alpha R^2$-gravity,
than for the classically equivalent Brans-Dicke theory. According to
our previous considerations, the modified Newtonian potential
in this case has the form
\beq
V(r) &=& - \,\frac{GM}{r}\,\Big(1 + \frac13\,e^{-m_0r} \Big)\,,
\label{Yuka}
\eeq
where the mass of the scalar mode $m_0$ can be very small only for a
huge value of the parameter $\alpha$. In the case when the enormous
value of $\alpha$ can overwhelmingly compensate the ``natural'' value
of  $m_0$  (which is of the Planck order of magnitude), the scalar
mass becomes incredibly small and the exponential in Eq.~(\ref{Yuka})
can be considered as constant unity at the astronomical scale. This
is exactly what we observe for the massless limit of the Brans-Dicke
model. In such an exotic situation one can not measure a real value
of the product $GM$ in laboratory experiments or in the Solar System
observations, and will observe  $(4/3)GM$ instead. At the same time
the bending of light will be measuring the real value $GM$, so some
discrepancy is unavoidable between the  two sets of observational
and experimental data.

In general, we will not bother with the redefinition of the product
$GM$, since we are not interested in such huge values of $\be$.
We will come back to this discussion only at one point, when
comparing the effect of the deflection of light to the modified
Newtonian potential.

For the sake of completeness we show explicitly how the scalar
contributions appear as a conformal transformation. Starting
from the auxiliary fields representation in
Eq.~\eqref{Coro_GeneralSolution}, if $\al = A = 0$ the general
solution of the field equations reads
\beq
h_{\mu\nu}^{(\alpha=A=0)}
&=&
h_{\mu\nu}^{(E)} + \Psi_{\mu\nu} + \bar{\Psi}_{\mu\nu}
- \frac{1}{2} \eta_{\mu\nu} \left(  \Psi + \bar{\Psi} \right) \,,
\eeq
which yields the metric
\beq
g_{\mu\nu}^{(\alpha=A=0)}
&=&
\eta_{\mu\nu} + \kappa \big[
h_{\mu\nu}^{(E)} + \Psi_{\mu\nu}
+ \bar{\Psi}_{\mu\nu}
-
\frac{1}{2} \eta_{\mu\nu}
\left(  \Psi + \bar{\Psi} \right) \big]\,.
\eeq
Thus, the metric associated to the full sixth-order gravity can
be expressed in the conformal form
\beq
g_{\mu\nu}
&=&
\left[ 1 - \kappa \left( \Phi + \bar{\Phi} \right)
+ \frac{\kappa}{2} \left(  \Psi + \bar{\Psi} \right) \right]
g_{\mu\nu}^{(\alpha=A=0)} ,
\eeq
keeping, as usual, terms up to first order in the metric fluctuation.

In what follows we consider systematically the results for the light
deflection according to the nature of the massive tensor excitations.
Namely,  we analyze the effective refractive index for the different
versions of $F_2$, as described in the previous section.

\subsection{Deflection with real simple poles}
\label{ss41}

In the case of real simple poles the effective refractive
index is given by the general formula
\beq
\label{Ref5}
n(r)
&=&
n_{\text{GR}}(r)
+ 2 MG \Bigg( \frac{m_{2+}^2}{m_{2-}^2
- m_{2+}^2} \frac{e^{-m_{2-}r}}{r}
- \frac{m_{2-}^2}{m_{2-}^2 - m_{2+}^2} \frac{e^{-m_{2+}r}}{r}\Bigg).
\eeq

Since $m_{2-} > m_{2+}$, the $m_{2-}$-term yields an attractive
force and produces an increase of $n(r)$, while the $m_{2+}$-term
gives a negative contribution to the refractive index, which is
responsible for the well-known repulsive force caused by the ghost
mode~\cite{Accioly15,modNew}. This repelling force is stronger
than that of the healthy massive mode, since
\beq
\label{tugofwar}
\bigg\vert
\frac{m_{2-}^2}{m_{2-}^2 - m_{2+}^2}
\frac{e^{-m_{2+}r}}{r} \bigg\vert
>
\bigg\vert \frac{m_{2+}^2}{m_{2+}^2 - m_{2-}^2}
\frac{e^{-m_{2-}r}}{r} \bigg\vert .
\eeq
As a consequence $\,n(r) < n_{\text{GR}}(r)$, implying that light
deflects less in the sixth-order gravity than in general relativity.

It is easy to show that for a fixed value of $\be$ there is also
the relation
\beq
n(r)>n_4(\beta,r) = n_{\text{GR}}(r) - \frac{2MG}{r}
\exp\Big( -\frac{4r}{|\beta|\kappa^2}\Big), \qquad
\label{4>6}
\eeq
where the {\it r.h.s.}
is the effective refractive index of the fourth-order gravity with the
same $\beta$, i.e., with $A = B = 0$. In order to prove inequality
\eqref{4>6}, we note that $\partial m_{2+}^2 / \partial B < 0$,
therefore the smallest value for $m_{2+}^2$ can be achieved by
taking the limit $\,B\to 0\,$ (remember $\,B < 0$), hence
\beq
\lim_{B\rightarrow 0} m_{2+}^2
&=&
-\, \frac{4}{\beta\kappa^2} \,=\, m_{2(4)}^2\,,
\eeq
which is precisely the square of the mass of the fourth-order
gravity's ghost~\cite{Stelle77}. Since the Yukawa potential is
stronger for a smaller mass, if $m_{2+}^2 = m_{2(4)}^2$
the repulsive term achieves its maximum strength, while the
attractive massive term tends to zero. We conclude that
$\,n(r) > n_4(r)\,$ for the same value of $\be$ in six- and
fourth-derivative models. In particular, $n(r) > 1$, which means
that the balance of the three forces never results in a net
outward deflection.

The previous discussion can be summarized by the following
chain of inequalities,  where the last two hold with the same
value of $\be$:
\beq
\label{InequalitiesClassic}
n_{\text{GR}}(r) \,>\, n(r) \,>\, n_4(r) \,>\, 1\,.
\eeq
Those inequalities become true equalities, respectively, in the
following limits:
\beq
&& {\it i)} \quad   m_{2\pm} \to \infty\,;
\nonumber
\\
&& {\it ii)} \quad   m_{2+}/m_{2-} \to 0
\,;\nonumber
\\
&& {\it iii)} \quad m_{2\pm} \to 0\,.
\label{cases}
\eeq
The possibility {\it ii)} corresponds to the fourth-derivative
gravity theory, with two disproportional masses as explained
in Sec.~\ref{sss311}, and with $B \rightarrow 0$ as we have
discussed above.

For the sake of completeness we write the result for the deflection
angle of a light ray with impact parameter $\rho$, given by
Eq.~\eqref{Angle} with the effective refractive index~\eqref{Ref5},
\beq
\label{integration}
\th
 &=&
 \th_{\text{GR}} \,+\, 2M G \rho \left(  I_- - I_+\right) ,
\eeq
\beq
\quad I_{\pm}
&=&
\dfrac{m_{2\mp}^2}{\vert m_{2\pm}^2 - m_{2\mp}^2\vert}
\,\int\limits_{-\infty}^{+\infty}
\Big(\dfrac{1}{r}  +   m_{2\pm}\Big)
\dfrac{e^{- r m_{2\pm}}}{r^2}\, dx\,,
\quad  \mbox{where}
\quad r = \sqrt{x^2+\rho^2}\,.
\label{def_r}
\eeq
Here $\theta_{\text{GR}} \equiv 4GM/\rho$ is the bending angle
predicted by general relativity.
In a higher derivative theory the ghost term
$I_{+}$ enters with a ``wrong'' sign, tending to reduce the
deflection angle.

The magnitude of deflection depends on the
three length scales, defined by the inverse masses of the tensor
modes and by the impact parameter. There is a region
delimited by $r_1 = 1/m_{2-}$ and $r_2 = 1/m_{2+}$
where the dominant contribution to the deflection is
owed to the ghost mode and the graviton.
If there is a strong hierarchy $\,m_{2-} \gg m_{2+}\,$ between the
masses, then the massive healthy tensor  mode is irrelevant along
the trajectory of the light ray, and the deflection angle is
approximately that of the fourth-order gravity~\cite{Accioly15},
\beq
\label{angle4order}
\theta
&\approx&
\theta_{\text{GR}}
- 2M G \rho \int\limits_{-\infty}^{+\infty}
\Big( \frac{1}{r} +  m_2\Big)
\frac{e^{-rm_2}}{r^2} \, dx\,.
\eeq
Here the definition of $r$ is the same as in (\ref{def_r}).
One can observe that outside of the sphere of the radius
$1/m_{2+}$, the dominant contribution to the light deflection
comes from the graviton sector and the effect of the massive
modes is suppressed.

\subsection{Deflection with real degenerate poles}
\label{ss42}

If the masses of the tensor excitations are approximately the same,
one can use the quantity $F_2$ given by
Eq.~\eqref{Fk_degenerate_eps}, or Eq.~\eqref{Fk_degenerate}
in the limit $m_{2-} = m_{2+} = m_2$. The latter yields the
effective refractive index
\beq
\label{index_degenerate}
n_{\text{degen}}(r)
&=&
n_{\text{GR}}(r) - 2 MG \left( \frac{1}{r}
+ \frac{m_2}{2} \right) e^{-m_2 r} \, .
\eeq
As far as the mentioned limit is smooth, it is possible to restrict our
consideration to the limit of equal masses. Without the hierarchy
between the masses, the relation~\eqref{tugofwar} and its implications do not hold.
 Then for a sufficiently small $r$ it is possible to have
 $n_{\text{degen}} < 0$. In this case the repulsive force is
 strong enough to cause a net outward deflection at this region.
 Hence, the
 chain of inequalities of Eq.~\eqref{InequalitiesClassic} simplifies
 to $n_{\text{GR}} > n_{\text{degen}}$, formally without a lower bound.

In this scenario, the expression for the deflection angle $\theta$
reads
\beq
\label{angle_degen}
\theta_{\text{degen}}
= \th_{\text{GR}} - 2M G \rho
\int\limits_{-\infty}^{+\infty}
\Big(
\frac{m_2^2\,r}{2} + m_2
+ \frac{1}{r}\Big)
\frac{e^{-rm_2}}{r^2} dx,
\eeq
with $r$ the same as in (\ref{def_r}),
which can be recognized as the deflection
angle in the fourth-derivative gravity \textit{with the same mass}
$m_2$ according to Eq.~\eqref{angle4order},  minus an extra
correction owed to the healthy massive (degenerate) excitation.
Indeed, the effective refractive index~\eqref{index_degenerate} can
be cast into  the form
\beq
n_{\text{degen}}(m_2,r) = n_4(m_2,r) - MG m_2 e^{-m_2 r}\,,
\label{dege}
\eeq
where $\,n_4(m_2,r)\,$ corresponds to fourth-order
gravity with the mass $\,m_2$.

It is important to stress that the strong repulsive force occurs only
at distances smaller than $1/m_2$. For the Planck-order mass
$m_2 \propto M_P$ this distance is of the order of $10^{-43}$~cm, so
 the repulsive effect does not affect the light deflected by
astronomical bodies including our Sun.

\subsection{Deflection with complex poles}
\label{ss43}

The expression for the effective refractive index of the sixth-order
gravity in the presence of complex massive poles follows from Eqs.~\eqref{m2_complex_potential} and~\eqref{Ref4},
\beq
n_{\text{C}}(r)
&=&
n_{\text{GR}}(r) \,-\,
2MG \left[ \cos(b_2 r) - \frac{\beta}{c_2}
\sin(b_2 r) \right]  \frac{e^{-a_2 r}}{r}\,.
\eeq
Accordingly, the deflection of a light ray with impact parameter
$\rho$ is given by
\beq
\th_{\text{C}}
&=& \th_{\text{GR}} \,-\,
2MG\rho \int\limits_{-\infty}^{+\infty}  dx
\Bigg\lbrace
\left[ b_2 - \frac{\beta}{c_2}
\left( a_2 + \frac{1}{r} \right) \right] \sin(b_2r)
\nonumber
\\
&+&
\left( a_2 + \frac{1}{r}  + \frac{\beta}{c_2} b_2 \right)
\cos(b_2r) \Bigg\rbrace
\,\frac{e^{-a_2r}}{r^2}\,,
\label{thetaComplexClassic}
\eeq
where we use the standard parametrization of (\ref{def_r}).

Since the expressions presented above for the deflection angles
carry the assumption that these angles are small, to all practical
purposes the impact parameter coincides with the closest approach
distance~\cite{FischbachFreeman}. Thus one can define the
trajectory scale by $\rho^{-1}$. In the complex poles cases there
are also three length scales: the one of the Yukawa part, the typical
length period of the oscillation and the impact parameter. The
analysis is complicated due to the presence of the oscillating
terms, hence in what  follows we describe only two simple but
illustrating examples.

\subsubsection{The case of $\,a_2 \gg b_2$}
\label{ss54}

From the definitions of Sec.~3.2, it follows that $a_2 > b_2$
if and only if $\beta < 0$. Besides, if $c_2$ is sufficiently small,
such that $\,c_2^2/\beta^2 \ll 1$, it is possible to have the real
part of the ``mass'' much larger than the imaginary part. In such
a scenario, the massive quantities  $a_2$ and $b_2$ may be
approximated by
\beq
a_2^2
& \approx &
\frac{4}{\kappa^2 \vert \beta \vert}
\left( 2 - \frac{3}{2} \frac{c_2^2}{\beta^2} \right)
\, , \qquad
b_2^2
\,\approx\,
\frac{2}{\kappa^2 \vert \beta \vert} \,\frac{c_2^2}{\beta^2}\,.
\eeq
Furthermore, the condition $a_2 \gg b_2$ means that the Yukawa
potential has a very short range if compared to the large space
period of the oscillatory terms.

It remains possible for the Yukawa and oscillation scales to
be either small or large with respect to the impact parameter.
Assuming that $\rho^{-1} \ll a_2$, the correction due to the
higher-derivatives is always tiny against the general relativity's
term, hence $\theta \approx \theta_{\text{GR}}$. The only
interesting situation is therefore $b_2 \ll \rho^{-1}$, with
$\rho^{-1}$ comparable to $a_2$. Accordingly we may write
$\cos(b_2r) \approx 1$ and $\sin(b_2r) \approx b_2r$, which
reduces the deflection angle to
\beq
\theta
& \approx &
\theta_{\text{GR}} - 2 MG \rho \int\limits_{-\infty}^{+\infty} dx\,
\left(  - \frac{\be a_2 b_2 }{c_2} \,r + a_2 + \frac{1}{r}   \right)
\frac{e^{-a_2r}}{r^2}
\nonumber
\\
 & \approx & \theta_{\text{GR}} - 2MG\rho
 \int\limits_{-\infty}^{+\infty} dx\,
 \left(  \frac{a_2^2}{2} \,r + a_2
 + \frac{1}{r}   \right) \frac{e^{-a_2r}}{r^2} \,.
\label{thetaComplexClassic4}
\eeq
It is easy to see that this is roughly the same
expression~\eqref{angle_degen} for the real degenerate
poles. This result should be expected, since the condition
$b_2 \ll \rho^{-1} \sim a_2$ means that the imaginary part
of $m_{2\pm}$ is tiny with respect to all other scales of the
system. Hence, to the leading order both scenarios turn out to
be the same, confirming the correctness of our calculations.
Differences start to emerge only when second- and
first-order corrections in $\,b_2r\,$ and $\,b_2/a_2$,
respectively, are taken into account.

\subsubsection{The case of  $b_2 \gg a_2$}
\label{sss55}

This condition only holds provided that $\be > 0$ and
$c_2^2 / \be^2 \ll 1$. The quantities $a_2$ and $b_2$ now
read, to the leading order,
\beq
a_2^2
&\approx &
\frac{2}{\kappa^2 \vert \beta \vert} \frac{c_2^2}{\beta^2}
\,, \qquad
b_2^2
\,\approx\,
\frac{4}{\kappa^2 \vert \beta \vert}
\left( 2 - \frac{3}{2} \frac{c_2^2}{\beta^2} \right) \,,
\eeq
and therefore $\,\,b_2^2/a_2^2 \approx 4 \beta^2 / c_2^2$.
As a consequence
\beq
\label{relationsScenario2}
b_2\frac{\vert\beta\vert}{c_2} \,\gg\,
b_2 \,>\, \frac{b_2}{2} \,\approx\, a_2\frac{\vert\beta\vert}{c_2}
\,\gg\, a_2\,.
\eeq
Let us remember that the condition $\,a_2 \ll b_2\,$ means that
the range of the Yukawa term is much larger than the space period
of the trigonometric functions in the expression for the effective
refractive index. Then many oscillations typically occur before the
exponential factor makes the whole expression negligible. This
regime, therefore, has a much stronger dependence on the impact
parameter if compared to the analysis which was described before.

In the regime $\rho^{-1} \gg b_2$ one may approximate the
argument of the trigonometric functions by the leading constant
value $\,\rho b_2$, for $r \approx \rho$, where the amplitude
of the correction term is maximum. It is clear that the change of
impact parameter by even a small fraction of its original value
can produce a large variation of the correction from the
higher-derivative terms, including altering the sign of
this correction.

This strong dependence on $\,\rho\,$ is only suppressed for
$\rho^{-1} < a_2$, due to the exponential damping. In view of
Eq.~\eqref{relationsScenario2}, the expression for the deflection
angle simplifies to
\beq
\label{correction2}
\theta
&\approx &
\theta_{\text{GR}} \,-\,
MG\rho \,\frac{\be b_2 }{c_2} \int\limits_{-\infty}^{+\infty}
dx \,\cos(b_2r)\,  \,   \frac{e^{-a_2r}}{r^2} \,.
\eeq

\subsection{Final comments on classical deflection}
\label{sss56}

Some general comments are in order. The two previous simple
examples show that the corrections due to the higher-derivative
terms can manifest strong dependence on the impact parameter in
the case of complex poles.
The origin of this effect is the oscillatory behaviour of the
effective refractive index. In the realistic situations, however,
the only feasible scenarios are those where the real part is large
enough to damp the oscillations far beyond the current
experimental bounds.
For instance, the most precise measurements of
deflection of light rays close to the Sun, carried out by
modelling solar occultations of radio sources, have confirmed
general relativity's prediction within the uncertainty of a few
parts in 100,000~\cite{Solar-radio}. In the visible spectrum, the
astrometry of stars during solar eclipses yield the verification
of the deflection angle to the precision of 1\%~\cite{Solar-visible}.

The deflection of light rays close to the solar limb in the
four-derivative gravity~\eqref{angle4order} corresponds to
the Yukawa potential with mass
$\,m_2 > 10^{-23}$~GeV~\cite{Accioly15}.
Such a figure, nevertheless, is far too small if one takes into
account laboratory tests of the inverse-square force law.
Torsion-balance experiments currently yield a much stricter
bound on the order of $\,m_2 >10^{-12}$~GeV for one
additional Yukawa potential~\cite{Kapner07,Giacchini16}.
These bounds may be viewed as first estimates to a lower-limit on
the real component $a_2$, if we assume that it is large enough to
damp the oscillations up to this length scale.
However, no bound on the imaginary part can be established
from this preliminary analysis. Precise modelling of experimental
data, especially those from torsion-balances, are required in order
to detect a possible oscillatory behaviour of the gravitational
potential. A stimulating discussion on the perspective of detecting
oscillations in the gravitational potential was set about in the
recent work~\cite{Perivolaropoulos}.

\section{Light bending: semiclassical approach}
\label{S5}

Let us now consider the photon as a quantum particle which interacts
with the classical external gravitational field. The main virtue of the
semiclassical calculation using Feynman diagrams is to consider the
background metric not as a completely sterile medium, but as an
external field whose massive modes are excited depending on the
energy of the interacting particle. In the case of the purely massless
gravitational excitation both classical and the semiclassical
approaches are equivalent, but in the presence of a massive parameter $m$ the
semiclassical scattering starts to depend on the ratio between $m$
and the energy of the photon~\cite{Accioly15,Caldwell}. As we
have already mentioned in the Introduction, the question of
whether and when the semiclassical approach can be used has been
discussed in the literature~\cite{Drummond&Hathrell} and the general
conclusion is that its pertinence is restricted to scattering
processes with very small impact parameter. Yet, this approach
looks interesting for it clarifies some general features of the
scattering, and therefore we include it here.
In what follows we present the results of the calculations for the
cross section formulas, and then discuss their applicability to
the bending of light in the Solar System.

\begin{figure} [t]
\centering
\includegraphics[scale=1]{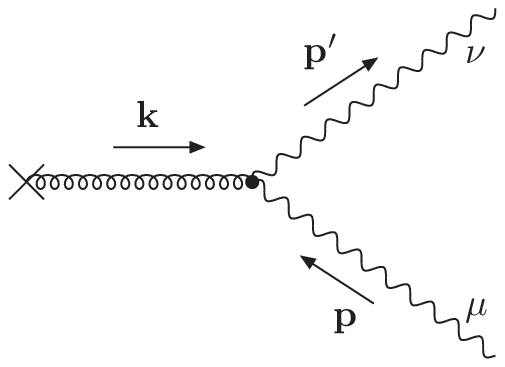}
\caption{Photon scattering by an external gravitational field.
Here $\vert\textbf{p}\vert = \vert\textbf{p}^\prime\vert$.}
\label{Fig1}
\end{figure}

At the tree level the only diagram contributing to the scattering
of a photon by a classical external gravitational field is the one
in Fig.~\ref{Fig1}, producing the vertex function
\beq
\label{vertice}
V_{\mu\nu}(p,p^\prime)
&=&
\frac{\ka}{2}\, \,h^{\la\rho}_{\text{ext}} (\textbf{k}) \,\,
F_{\mu\nu\la\rho} (p,p^\prime) \,,
\eeq
where
$\,p\,$ and $\,p^\prime\,$ are the four-momenta of the initial and
final states of the photon, while
$\,h^{\la\rho}_{\text{ext}} (\textbf{k})\,$ is the linearized
gravitational  field in the momentum-space representation.
The function in~\eqref{vertice}  has the form
\beq
F_{\mu\nu\la\rho} (p,p^\prime)
&=&
-\,\eta_{\mu\nu}\eta_{\la\rho} p\cdot p^\prime
\,+\, \eta_{\lambda\rho}p^\prime_\mu p_\nu
+  2 \big( \eta_{\mu\nu}p_\la p_\rho^\prime
- \eta_{\nu\rho} p_\la p_\mu^\prime
- \eta_{\mu\la}p_\nu p_\rho^\prime
\nonumber
\\
&& + \, \eta_{\mu\la}\eta_{\nu\rho}
p\cdot p^\prime \big) .
\label{FVertex}
\eeq

Since, according to the Theorem~\ref{Theorem1}, the
gravitational field $\,h^{\la\rho}_{\text{ext}}\,$
can be written as the sum of five auxiliary fields,
and owed to the linearity of the Fourier transform,
the vertex function assumes the form
\beq
\label{amplitudes}
V_{\mu\nu}
&=&
\mathcal{M}_{\mu\nu}^{(E)}  + \mathcal{M}_{\mu\nu}^{(\Psi)}
+ \mathcal{M}_{\mu\nu}^{(\bar{\Psi})}
+ \mathcal{M}_{\mu\nu}^{(\Phi)}
+ \mathcal{M}_{\mu\nu}^{(\bar{\Phi})}\,,
\\
\nonumber
\mathcal{M}_{\mu\nu}^{(\dots)}
 &=&
\frac{\ka}{2}\, h^{(\dots)\lambda\rho}_{\text{ext}}\,\,
F_{\mu\nu\lambda\rho}\,,
\eeq
where the last equation is valid for all five auxiliary fields.

We point out that for a photon the dispersion relation is
$\,p^2 = E^2 - \textbf{p}^2 = 0 = {p^{\prime}}^2$.
Furthermore, we can assume the field to be weak and
hence neglect the possible energy exchange between the photon
and gravitational field. Therefore it follows that
$\,\vert\textbf{p}\vert = \vert\textbf{p}^\prime\vert$.
Bearing this in mind, it is easy to verify that
$\,\eta^{\lambda\rho} F_{\mu\nu\lambda\rho} = 0$.
Hence
$\,\mathcal{M}_{\mu\nu}^{(\Phi)}
= \mathcal{M}_{\mu\nu}^{(\bar{\Phi})} = 0\,$ and the
Feynman amplitudes related to the scalar modes of the
gravitational field are null.

From the perspective of Feynman diagrams, the
contribution of the scalar mode of the metric vanishes
because it interacts with the photon through the
trace of the energy-momentum tensor. This trace is
null for the electromagnetic field, and as a consequence
{\it none} of the scalar components contribute to the
scattering of light. This confirms the result which we obtained
in Sec.~\ref{S4} within the classical framework. One of the
manifestations of this is that the $R^2$-  and
$R\square R$-terms do not affect light deflection,
except in the recalibration of the product $GM$ in
the special case of very light scalar mode(s).

The Feynman amplitude for the scattering of photons with
initial polarization vector $\ep^\mu_r(\textbf{p})$ and
final  polarization $\ep^\nu_{r^\prime}(\textbf{p}^\prime)$
is given by
\beq
\mathcal{M}_{rr^\prime} = V_{\mu\nu}(p,p^\prime) \,\ep_r^\mu(\textbf{p})
\,\ep_{r^\prime}^\nu(\textbf{p}^\prime)\,.
\eeq
Taking into account Eq.~\eqref{FVertex} and
the completeness relation for the polarization vectors,
\beq
\label{Polarization}
\sum_{r=1}^2
\ep^\mu_r(\textbf{p}) \,\ep^\nu_r(\textbf{p})
&=&
- \,\eta^{\mu\nu} - \frac{p^\mu p^\nu}{(p\cdot n)^2}
+ \dfrac{p^\mu n^\nu + p^\nu n^\mu}{p\cdot n}
\,,\quad
(n^\mu n_\mu=1).
\nonumber
\eeq
The sum over all the polarizations yields the unpolarized
cross section
\beq
\label{decomposicaoSecaoChoque}
\dfrac{d\sigma}{d\Omega} \,=\, \frac{1}{2(4\pi)^2}
\sum_{r,r^\prime} \vert \mathcal{M}_{rr^\prime} \vert^2
= \frac{1}{2(4\pi)^2} V_{\mu\nu} V^{\mu\nu} .
\eeq
Furthermore, it is cursory to show that
\beq
\label{Identidade}
\eta^{\la 0}\eta^{\rho 0}
\,F_{\la\rho\mu\nu}\,\eta_{\al 0}\eta_{\be 0}
\,F^{\al\be\mu\nu}
&=&
2 E^4 (1-\cos \theta)^2\,,
\eeq
where $E=E^\prime$ is the energy of the photon and $\th$ is
the deflection angle between $\textbf{p}$ and $\textbf{p}^\prime$.

Using Eqs.~\eqref{amplitudes},
\eqref{decomposicaoSecaoChoque} and \eqref{Identidade} it follows that
\beq
\label{CrossSection-com}
\dfrac{d\sigma}{d\Omega}
&=&
\dfrac{\kappa^4 M^2 E^4 (1+\cos\theta)^2}{(16\pi)^2}
\Bigg[ \frac{1}{\textbf{k}^2} - \frac{1}{m_{2-}^2 - m_{2+}^2}
 \left( \dfrac{m_{2-}^2}{\textbf{k}^2 + m_{2+}^2}
- \dfrac{m_{2+}^2}{\textbf{k}^2 + m_{2-}^2} \right)  \Bigg] ^2.
\eeq
In the formula~\eqref{CrossSection-com} one can recognize the
standard gravitational version of the Rutherford formula, plus
the correction coming from the massive modes.

In what follows we assume that the bending angle is small
and calculations are performed in the leading order in $\theta$.
Then
$\,{\vec k}^2 \approx 2 {\vec p}^2 (1-\cos\th) \approx E^2\th^2\,$
and the previous expression reduces to
\beq
\label{CrossSection}
\dfrac{d\sigma}{d\Omega}
&=&
16 G^2 M^2 \Big(\dfrac{1}{\theta^2}
- \dfrac{m_{2-}^2}{m_{2-}^2 - m_{2+}^2}
\dfrac{E^2}{E^2\theta^2
+ m_{2+}^2}
+ \dfrac{m_{2+}^2}{m_{2-}^2 - m_{2+}^2}
\dfrac{E^2}{E^2\theta^2 + m_{2-}^2} \Big)^2 ,
\eeq
where we omitted $\,{\cal O}\big(\th^{-3}\big)\,$ and other
relatively small terms.

It is clear that the propagation of photons in this model is dispersive,
i.e., depends on the energy of the photon. The same general
feature was established in Ref.~\cite{Accioly15} for the photons in
the fourth-order theory. However, in the six-derivative case there are
several possible scenarios, depending on the type of the quantities
$m_{2\pm}$. In what follows we treat each case separately.

\subsection{Scattering with real simple poles}
\label{ss51}

Let us start by recalling that in the simpler fourth-order
gravity, for a given $\theta$, the cross section is smaller
than in general relativity,
\beq
\left( \frac{d\sigma}{d\Omega} \right)_4
&=&
16 G^2 M^2 \left( \frac{1}{\theta^2} - \dfrac{E^2}{E^2\theta^2
+ m_{2}^2} \right)^2
\nonumber
\\
&<&
\frac{16 G^2 M^2}{\theta^4}
\,=\,
\left( \frac{d\sigma}{d\Omega} \right)_{\text{GR}}.
\eeq
This happens because the $R_{\mu\nu}^2$-sector yields a repulsive
dispersive interaction such that more energetic photons are less
scattered.

%

In the sixth-order gravity, in addition to the attractive non-dispersive
force coming from the $R$-sector and the repulsive dispersive force
due to the $R_{\mu\nu}^2$-sector, there is another attractive,
dispersive, force due to the term $R_{\mu\nu}\square R^{\mu\nu}$.
This makes the ``tug of war'' between those forces more complicated
than in the fourth-order gravity; yet, the qualitative conclusions are
the same. One can summarize the results as follows:
\begin{itemize}
\item[i.] Light is less scattered than in general relativity.
The hierarchy $m_{2-} > m_{2+}$ implies that
\beq
\frac{m_{2-}^2 E^2}{E^2\theta^2 + m_{2+}^2}
\,>\,
\frac{m_{2+}^2 E^2}{E^2\theta^2 + m_{2-}^2}\,,
\label{103}
\eeq
and hence (see Eq.~\eqref{CrossSection})
\beq
\label{InequalitiesSemiClassic}
\left( \frac{d\sigma}{d\Omega} \right)_{\text{GR}}
\,>\,
\frac{d\sigma}{d\Omega}
\,>\,
\left( \frac{d\sigma}{d\Omega} \right)_4 > 0\,,
\eeq
where $\left( \frac{d\sigma}{d\Omega} \right)_4$ is the
cross section  for the fourth-order gravity with the same $\be$
(see discussion in the Sec.~\ref{ss41}). The second inequality
tends to equality in the case of strong hierarchy
$\,m_{2-} \gg m_{2+}$, while the last inequality in~\eqref{InequalitiesSemiClassic}
tends to equality in the limit $E \to \infty$, when no deflection
occurs.

\item[ii.]

More energetic photons undergo less deflection. This happens
because they interact strongly with
the dispersive terms and, as
one can see in~\eqref{103}, among the dispersive forces the
repelling one is always bigger. Physically, the reason is that the
coupling constant is the same for all intermediate tensor bosons,
thus the one with larger mass makes smaller effect.

\end{itemize}

The dependence on $E$ cannot be observed in the classical
approach. But it is interesting to note that besides the dispersive
behaviour, the general qualitative conclusions of the classical approach
are verified at the quantum level. In order to see this, one can
compare, for instance, the chain of inequalities in
Eqs.~\eqref{InequalitiesClassic} and~\eqref{InequalitiesSemiClassic}.

\subsection{Scattering with real degenerate poles}
\label{ss52}

The cross section for the case of real degenerate poles can be
explored using the general expression for the cross section
\eqref{CrossSection}.  We start from the case of a weak hierarchy
\beq
m_{2-} = m_{2+} + \ep \,=\, m_2 + \ep,
\quad \mbox{with}
\quad
\frac{\ep}{m_2} \ll 1,
\nonumber
\eeq
%
and then take the limit $\ep \to 0$, which smoothly yields
\beq
\label{CrossSection_degenerate}
\left( \frac{d\sigma}{d\Omega} \right)_\text{degen}
&=&
16 G^2 M^2 \bigg[ \frac{1}{\th^2} - \frac{E^4\theta^2 + 2m_2^2E^2}{(E^2\th^2 + m_2^2)^2}
\bigg]^2 .
\eeq
It is straightforward to verify that this cross section is bounded
by zero (for $m_2/E \to 0$) and by the general relativity cross
section as $\,E/m_2 \to 0$. Therefore the qualitative conclusions
of the case with real simple poles apply here too; namely, light
deflects less than in general relativity, and more energetic
photons are less scattered.

\subsection{Scattering with complex poles}
\label{ss53}

The unpolarized cross section for the situation where the poles of
the propagator are complex can be evaluated by inserting the
quantities \eqref{complexmassM2} and~\eqref{a2_and_c2_def}
into the general formula~\eqref{CrossSection}. This procedure yields
\beq
\label{CrossSectionComplex}
\left( \dfrac{d\sigma}{d\Omega} \right)_{\text{C}}
\,=\,
16 G^2 M^2 \left( \frac{1}{\th^2} - f \right)^2,
\eeq
where
\beq
f
&=&
\dfrac{E^4\theta^2 + 2E^2(a_2^2 - b_2^2)}{\left( E^2\theta^2
+ a_2^2 - b_2^2\right) ^2 + 4 a_2^2 b_2^2} \, .
\label{f}
\eeq

Differently from the case of real poles, for certain angles and
combinations  of $\,B, \beta$ and $E$ it is possible to have
$\,\left( \frac{d\sigma}{d\Omega} \right)_{\text{C}}
\geq \left( \frac{d\sigma}{d\Omega}\right)_{\text{GR}}$.
A useful example is as follows:
\beq
\label{exampleComplexSemi}
\th^2 = \dfrac{\beta \pm \sqrt{\beta^2
+ \frac{8B}{\kappa^2}}}{2 \vert B\vert E^2}
\, \Longrightarrow \,
\left( \frac{d\sigma}{d\Omega}\right)_\text{C} (\th)
= 4 \left( \frac{d\sigma}{d\Omega}\right)_{\text{GR}} (\th) \,.
\eeq
It is good to remember that
Eq.~\eqref{exampleComplexSemi} only holds if
$\,\be > 0$, otherwise $\,\th^2 < 0$.

It is natural to ask whether it is possible to have
$\,\left( \frac{d\sigma}{d\Omega}\right)_\text{C}
> \left( \frac{d\sigma}{d\Omega}\right)_{\text{GR}}\,$
with $\,\be < 0$. In order to answer this question, we must note
that the quantity $\,f\,$ which appears on the cross section
\eqref{CrossSectionComplex} is always positive if
$\,a_2 > b_2$, but has indefinite sign if $\,a_2 < b_2$. In
view of this fact, we analyse each possibility separately, as
well as the special case $a_2 = b_2$.

\subsubsection{The case of $a_2 > b_2$}

It is straightforward to verify that $f$ in Eq.~\eqref{f} is not only
positive, but is also a strictly increasing function on $E$, if
$a_2 > b_2$ (or, equivalently, $\be < 0$). In fact, the sign
of $\,\pa f/\pa E\,$ is determined by its numerator,
\beq
\sign \left( \frac{\partial f}{\partial E}\right)
=
\sign
\Big[
4 a_2^2 b_2^2 (a_2^2 - b_2^2) + 4 E^2 \theta^2 a_2^2 b_2^2
+ (a_2^2 - b_2^2)^3 + E^2 \theta^2 (a_2^2 - b_2^2)^2 \Big]\,.
\eeq
Hence, if $\beta < 0$, the function $f$ grows with the increase of
$E$. Besides, $\,f \to 1/\th^2$  when $E \to \infty$, which means that
sending photons with higher energy can, at most, cancel the
Einstein's term $1/\th^2$ in the cross section expression. We
conclude that if $\be < 0$ then light would always scatter less
than in general relativity, and even less for high-energy photons.
This is qualitatively the same behaviour as in the case of real poles.

In the strong hierarchy regime $a_2 \gg b_2$, the cross section formula~\eqref{CrossSectionComplex} boils down to
\beq
\dfrac{d\sigma}{d\Omega}
&\approx &
16 G^2 M^2 \left[ \frac{1}{\th^2}
- \dfrac{E^4\theta^2 + 2 a_2^2 E^2}
{\left( E^2\theta^2 + a_2^2\right) ^2} \right] ^2\,.
\eeq
As one ought to expect, this expression corresponds
to the cross section for real degenerate
poles~\eqref{CrossSection_degenerate}. Indeed, in
Sec.~\ref{ss54} it was argued that both situations are
equivalent if terms of order $b_2/a_2$ are not taken into account.

\subsubsection{The case of $a_2 < b_2$}

In the six-derivative theory one can set $\be > 0$ and still have a
stable massless tensor mode. Then $a_2 < b_2$, hence it is
possible to have $f < 0$ and $\frac{\partial f}{\partial E} <0$,
according to the conditions
\beq
f < 0 \,\, \Longleftrightarrow \,\,
b_2^2 - a_2^2 \,>\, \frac{E^2\theta^2}{2},
\\
\frac{\partial f}{\partial E} \,\,<\,\, 0
\,\,  \Longleftrightarrow \,b_2^2 - a_2^2
\,\,> \,\,E^2\theta^2 \,.
\eeq
It is easy to see that the two following regimes may occur, in
addition to the usual behaviour of the previously described
scenario. First, if $\,f < 0\,$ but $\,\frac{\pa f}{\pa E} > 0$,
then the correction term $\,f\,$ will sum up with the general
relativity term $1/\theta^2$, making the cross section larger
than the general relativity one. At the same time more energetic
photons still have smaller cross section. For low energy photons,
the cross section increases with the energy up to the point where
$\,E^2\th^2 = b_2^2 - a_2^2$.  Below this value of energy the
sign of  derivative changes and the cross section starts to decrease.

The zero point of the derivative $\, \pa f / \pa E\,$
corresponds to the unique local minimum of $\,f(E)$. For lower
energy photons both $\,f<0\,$ and $\,\frac{\pa f}{\pa E} < 0$,
hence the cross section is still greater than in general relativity,
but it decreases to
$\,\left( \frac{d\si}{d\Om}\right)_{\text{GR}}\,$ as
$\,E \to 0$. At this region more energetic
photons undergo more scattering. Therefore, $f$ is bounded
between $\,-\frac{E^4\theta^2}{4a_2^2b_2^2}\,$
and $\,\th^{-2}$, and the cross section satisfies the conditions
\beq
0 \,\leq \,\dfrac{d\sigma}{d\Omega} \,\leq \,
16 MG \left[ \frac{1}{\th^2}
+ \frac{E^4\th^2}{4a_2^2b_2^2} \right] ^2\,.
\eeq

One can note that if the massive parameters of the action are
of the order of the Planck mass, then the upper bound on the
cross section is going to be very close to the cross
section of general relativity. Hence, this scenario is not ruled
out in principle. It is interesting to notice that this is the only
scenario where the upper-bound on the cross
section is not trivial.

\subsubsection{The case $a_2 \approx b_2$}

The condition
$\,a_2 \approx b_2 = \mu\,$ is fulfilled provided that
$\,\ka^2 \be^2 \ll 16 \vert B \vert$. Under such an assumption the
massive parameter reads $\mu \approx \left( \ka^2 |B|\right) ^{-1/4}$,
and the cross section becomes
\beq
\dfrac{d\sigma}{d\Omega}
\,\approx\,
16 G^2 M^2 \left[ \frac{1}{\theta^2}
- \dfrac{E^4\theta^2}{4 \mu^4 + E^4\theta^4 } \right] ^2
\,\leq\, \left( \dfrac{d\sigma}{d\Omega}\right)_{\text{GR}}.
\eeq

As expected,
$\,\frac{d\si}{d\Om} \to \left( \frac{d\si}{d\Om}\right)_{\text{GR}}\,$ when
$\,\mu/E \to \infty\,$. Hence, as in the scenario with
real poles or with $a_2 > b_2$, the cross section decreases with
the energy of the photon.

\subsection{On the applicability of semiclassical approach}

Let us now comment on the applicability of the diagrammatic
approach for the gravitational light bending, which has been
described in this section. It is well known that this method is
equivalent to the classical one for evaluation of the modified
Newtonian potential in both general relativity and higher
derivative gravity. This approach also works pretty well in
general relativity for the description of the bending of light.
At the same time, we know that for the fourth-derivative gravity
the results of the classical and semiclassical methods diverge
\cite{Accioly15}, and we have just seen that the situation is
the same in the six-derivative gravity case. Therefore, it is
necessary to explain the discrepancy between the two methods
and understand which of them is correct and which is not.

The semiclassical approach implies the evaluation of the scattering
amplitude, representing the interaction of a photon with a massive
matter source. It is usually assumed that this massive particle is
heavy and remains static, since it represents a heavy body such as
a star or a galaxy, while the photon plays the role of a test particle.
At the tree level this corresponds to a Feynman diagram as displayed
in Fig.~\ref{Fig1}. In the case of general relativity, the
cross section for the exchange of one graviton is simply a reduced
case of Eq.~\eqref{CrossSection-com},
\beq
\label{CrossSection-GR}
\left( \dfrac{d\sigma}{d\Omega} \right)_{GR}
&=&
\,\,
\frac{\kappa^4 M^2}{(8\pi)^2} \dfrac{E^4}{\textbf{k}^4},
\eeq
which in the small-angle approximation boils down to
\beq
\left( \frac{d\sigma}{d\Omega} \right)_{GR}
&=&
\frac{16 G^2 M^2}{\th^4}.
\eeq
This matches the small-angle classical cross section for general
relativity~\cite{NuoCim67}, but it is not a trivial fact. It only
happens because of the special form of the interaction, which
has an infinite range or, in other words, does not have an intrinsic
scale~\cite{Drummond&Hathrell}. This interaction classically corresponds
to the Newtonian potential, and its remarkable
feature is that the classical, the Born-approximated and
the exact quantum cross sections do coincide~\cite{Bohm_book}.

In the very simple terms we can understand the validity of the
semiclassical approximation in this case as follows. The underlying 
assumption in the quantum formulation is that the initial and final 
states of the photon are described by a wave which has no space
localization. Therefore, the massless intermediate
particle provide a non-scale description, such that the absence
of localization of the free photon in the quantum formalism
does not manifest as a trouble in the calculations.
However, if the same scheme is applied to
the theory with massive intermediate particles, the result
may be incorrect, especially if the range of the force
is  not much larger than the impact parameter of the
given scattering process. In other words, in the case of a massive
intermediate particle one cannot regard the initial and
final states of the scattered particle as a free wave without
space localization, unless the impact parameter is sufficiently
small.

Let us consider the issue in more detail. In order to apply quantum
cross sections for evaluating the deflection of a photon passing
close to an astronomical body, one has to compare quantum and
classical cross sections~\cite{Berends&Gastmans,Delbourgo&Phocas,
Caldwell,HugginsToms87,Accioly15,Accioly&al},
\beq
\label{Semiclassical}
\frac{b}{\sin\th} \,\bigg\vert \frac{db}{d\theta} \bigg\vert
&=&
\left( \frac{d\sigma}{d\Omega}\right)_{\text{quantum}}.
\eeq
Solving the differential equation we arrive at the answer for
$\theta$ as a function of the impact parameter $b$.
As we have already noted above, the semiclassical methodology
based on (\ref{Semiclassical}) cannot be safely used in most of
the cases, since it gives the correct result only in the case of
tree-level general relativity \cite{Drummond&Hathrell} (see
also more recent discussion in~\cite{Holstein}).

The reason for the failure of using (\ref{Semiclassical}) is related
to the fundamental difference of the terms on both sides of this
equation. In the classical scattering theory there is a direct
relation between the impact parameter and the scattered angle.
At the same time the quantum cross section has an intrinsic
probabilistic meaning, for it is related to the amplitude of the
scattered wave function. It assumes that the incoming particle
can be well represented by a plane wave, and that the scattered
particle is going to be detected far away from the interaction
zone. Such assumptions should not be taken for granted in all
cases.

Consider as an example the sixth-order gravity with real poles.
Following the extremely ``mild'' assumption which was already
used in in Sec.~\ref{S4}, let us assume that the masses
$m_{2\pm}$ are such that the Yukawa potentials have ranges
on the submillimeter scale, in agreement to the lower bounds from
 laboratory experiments~\cite{Kapner07,Giacchini16}. Then, classically,
a light ray with impact parameter of one solar radius $R_\odot$
would undergo roughly the same deflection as in general relativity.
On the other hand, the quantum cross section depends on the
energy of the photon, and can become arbitrarily small provided
that $\,E \gg m_{2\pm}$. The last means that no appreciable
deflection should occur if the wavelength of the photon is
short enough, e.g., at the submicrometer scale.

The contradiction occurs because in the present case it is not
correct to use the quantum cross section as for large
impact parameters we are not in the quantum regime. The
light emitted by a distant star and deflected by the Sun
with the massive intermediate particle cannot be represented
by a probabilistic plane wave interacting with the corresponding
Yukawa potential. Instead, it ought to be described by a compact
wave packet arriving with a definite impact parameter
$b \sim R_\odot$  and hence it is passing far away from
the centre of the potential.  The quantum cross section should
be used only when the impact parameter is comparable to
the size of the wave packet. The typical scale involved in the
problem of our interest is the one defined by the massive tensor
modes. It is clear that this condition cannot be achieved at the
macroscopic astronomical scales.

The correct way to use the tree-level scattering amplitudes
for evaluating the gravitational bending of light by astronomical
bodies is via its Fourier transform, which provides the classical
interaction potential. This quantity may be used within the
classical scattering theory to compute the bending angle. The
methodology is equivalent to the classical analysis presented
in Sec.~\ref{S4} and agrees with the common lore that the
tree-level computations, in general, should agree with the
classical physics results.

\section{Conclusions}
\label{S6}

The six-derivative model represents the simplest version of the large
class of quantum gravity theories which are local (i.e., polynomial
in derivatives), superrenormalizable and that enable one to have only
complex conjugate pairs of massive poles in the propagator.
According to the recent paper~\cite{LM-Sh} this kind of theories
have unitary $S$-matrix and therefore resolve an old-standing
conflict between renormalizability and unitarity in quantum gravity.
Another class of theories which possess similar properties are
non-local, (or non-polynomial in derivatives) and have no massive
poles at the tree level
\cite{krasnikov,kuzmin,Tseytlin-95,Tomboulis-97}.
However, in these theories an infinite number of ghost-like states
with complex poles emerge when any kind of quantum loop corrections
are taken into account~\cite{CountGhost}.  For this reason the
theory with higher derivatives and complex massive poles is quite
general in quantum gravity, and therefore it deserves serious
investigation not only in the UV, but also in the IR limit.

In the present work we made the first step in exploring the
low-energy manifestations of complex higher-derivative states.
For the sake of completeness we also considered the cases
of real massive poles, both simple and multiple. It turned out
that the effect of complex poles on the modified Newtonian
potential and on the gravitational bending of light is partially
similar to the one of the massive real ghost mode in the
four-derivative gravity theory. At the same time, there are
some new and remarkable features, such as the oscillatory
behaviour of the potential $V(r)$, which takes place in the
case of complex poles.

We have shown that there is a difference between the classical and
quantum cross sections for the gravitational scattering of the
photon. In the former case one has to treat photon as a particle
moving in the determined classical background of a weak gravitational
field. Contrary to this, within the semiclassical approach the tree-level
cross section is used to evaluate the same scattering.
We have confirmed, for the six-derivative models, the previous
conclusions of~\cite{Drummond&Hathrell}, that the semiclassical
approach used, e.g., in~\cite{Berends&Gastmans} cannot be
applied for higher-derivative models, except in the case of extremely
small impact parameters.

Still in the quantum domain, it was shown that
the cross sections vanish for photons with energies $E \gg |m_{2\pm}|$.
This feature was also noticed in the case of fourth-order gravity~\cite{Accioly15}
and it can be qualitatively explained recalling the uncertainty principle.
In fact, this is the energy
necessary to localize a particle with uncertainty smaller than
$\,1/|m_{2\pm}|$. In such case the Coulomb-shielding property
of the Yukawa short-range potentials is broken, and the photon
is able to probe the inner parts of the potential, where it tends
to behave like $1/r$. As the contribution
of the Yukawa-type potentials approaches that of Rutherford
scattering, they cancel out the authentic Rutherford term
owed to the (massless) graviton. From the diagrammatic
perspective, this can be understood as the back-reaction of the
photon on the background and its capability of exciting the massive
modes. The effect becomes significant for high-energy photons with
small impact parameters, with frequencies comparable to the mass
of the tensor excitations.

From the phenomenological side, our investigation has shown that
the gravitational light bending in the Solar System cannot predict
new dispersive phenomena such as in lensing or arriving time delays,
nor give tight constraints to the massive modes. It is more likely to
detect the influence of the higher-derivative terms in laboratory
experiments using torsion-balance or in the cosmological observations.
The analysis of these possibilities would be quite interesting and
should represent an interesting subject for future work.

\appendix
\section{Proof of the Theorem~\ref{Theorem1}}
\label{S7}

Let us prove Theorem~\ref{Theorem1} which enables one to write the general
of the field equations in terms of auxiliary fields.

It is easy to show that the gauge condition $\Ga_\mu = 0$
can be achieved by means of coordinate transformation
$x^\mu \to x^{\prime\mu} = x^\mu + \kappa \xi^\mu(x)$.
The transformation of the linearized perturbations are
\beq
h^\prime_{\mu\nu} &=& h_{\mu\nu} - (\xi_{\mu,\nu}
+ \xi_{\nu,\mu}) ,
\\
\ga^\prime_{\mu\nu} &=&
\ga_{\mu\nu} - \left( {\xi}_{\mu,\nu} + \xi_{\nu,\mu} \right) +\eta_{\mu\nu}{\xi_\lambda}^{,\lambda}
 \,.
 \eeq
Since for the scalar curvature $R^\prime = R$, it is easy to derive
\beq
\Ga_\mu
& \rightarrow & \Ga^\prime_\mu
=
\Ga_\mu
- \left( 1 - \frac{\ka^2\be}{4} \square
- \frac{\ka^2 B}{4} \square^2 \right) \square \xi_\mu .
\eeq
The next step consists in the following proposition:

\begin{propo}
The general solution of the system
\beq
\label{SystemAppendix}
&&
\left( 1 - \dfrac{\kappa^2\beta}{4}\square
- \dfrac{\kappa^2 B}{4}\square^2\right)
\left( - \square h_{\mu\nu}
+ \dfrac{R}{3\kappa} \eta_{\mu\nu} \right)
\,=\,
\dfrac{\kappa}{2} \left( T_{\mu\nu} - \dfrac{T}{3} \eta_{\mu\nu}\right) \,,
\\
&& \Gamma_{\mu} \,=\,
\left( 1 - \dfrac{\kappa^2\beta}{4}\square
- \dfrac{\kappa^2 B}{4}\square^2 \right) {\gamma_{\mu\rho}}^{,\rho}
- \dfrac{\kappa}{2}
\left( \alpha + \dfrac{\beta}{2} + A \square
+ \dfrac{B}{2} \square \right) R_{,\mu} = 0. \label{SystemBAppendix}
\eeq
has the form
\beq
\label{Theorem_GeneralSolution}
h_{\mu\nu}
&=&
h_{\mu\nu}^{(E)} + ( m_{2+}^2 + m_{2-}^2 + \square )
\psi_{\mu\nu}
 - \, \eta_{\mu\nu} ( m_{0+}^2 + m_{0-}^2
+ \square ) \phi\,,
\eeq
where the fields $\,h_{\mu\nu}^{(E)}$,
$\,\psi_{\mu\nu}\,$ and $\,\phi\,$ satisfy the equations
\beq
\label{Theorem_h(E)}
&&
\square h_{\mu\nu}^{(E)}
\,=\,
\dfrac{\kappa}{2} \left( \dfrac{1}{2}\,T \eta_{\mu\nu}
- T_{\mu\nu} \right) ,
\label{hE_1}
\\
&&
{\gamma_{\mu\nu}^{(E),\nu}} \,=\, 0 ,
\quad
\mbox{\rm where}
\quad
\gamma_{\mu\nu}^{(E)} \,=\, h_{\mu\nu}^{(E)}
- \dfrac{1}{2} \eta_{\mu\nu} h^{(E)},
\label{hE_2}
\\
\label{Theorem_Psi}
&&
( m_{2+}^2 + \square ) ( m_{2-}^2 + \square ) \psi_{\mu\nu}
\,=\,
\dfrac{\kappa}{2} \left( T_{\mu\nu}
- \dfrac{1}{3}\,T \eta_{\mu\nu} \right) ,
\nonumber
\\
\label{psi_1}
\\
&&
(m_{2+}^2 + m_{2-}^2 + \square) ({\psi_{\mu\nu}}^{,\mu\nu}
- \square \psi) = 0,
\label{psi_2}
\\
\label{Theorem_Phi}
&&
( m_{0+}^2 + \square ) ( m_{0-}^2 + \square ) \phi
= \dfrac{\kappa T}{12}.
\eeq
\end{propo}
\noindent
Here we used notations~\eqref{sigmas} and~\eqref{Def_masses}.

\textit{Proof:}
The first parenthesis in Eq.~\eqref{SystemAppendix} can be
factorized as
\beq
-\,\frac{\ka^2 B}{4}\,(m_{2+}^2 + \square)(m_{2-}^2 + \square)\,,
\eeq
provided that
\beq
m_{2+}^2 + m_{2-}^2 = \frac{\beta}{B}
\,
\text{ and }
\,
m_{2+}^2 m_{2-}^2 = - \frac{4}{\kappa^2 B},
\eeq
that corresponds to the definition~\eqref{Def_masses}. Defining
\beq
\psi_{\mu\nu} \,=\, - \frac{\kappa^2 B}{4} \left( - \square h_{\mu\nu} + \dfrac{1}{3\kappa} R \eta_{\mu\nu} \right) ,
\eeq
Eq.~\eqref{SystemAppendix} results in
\beq
\label{eq_mot_psi}
(m_{2+}^2 + \square)\,(m_{2-}^2 + \square) \,\psi_{\mu\nu}
 \,=\, \dfrac{\kappa}{2} \left( T_{\mu\nu}
 - \dfrac{1}{3}\,T \eta_{\mu\nu}\right) ,
\eeq
which is precisely \eqref{psi_1}. In terms of the field
$\psi_{\mu\nu}$, Eq.~\eqref{SystemAppendix} can be rewritten as
\beq
\label{deriv_hE}
\square^2 \psi_{\mu\nu} + \frac{\beta}{B} \square \psi_{\mu\nu}
- \square h_{\mu\nu} + \frac{R}{3\kappa} \eta_{\mu\nu}
=
\frac{\kappa}{2} \left( T_{\mu\nu}
- \frac{1}{3}\,T \eta_{\mu\nu} \right) .
\eeq

This equation can be cast in a more useful form by means of
the following expressions:

\begin{itemize}
\item[i)] Trace of \eqref{SystemAppendix},
\beq
\left[ 1 - \frac{\kappa^2}{4}
\left( \beta  + B \square \right)\square  \right]
\left( \square h - \frac{4}{3\kappa} R \right)
\,=\, \frac{\kappa}{6}\,T\, .
\eeq
\item[ii)] Divergence of $\,\Ga_\mu\,$ in \eqref{SystemBAppendix}
\beq
 0
=
\left[ 1 - \frac{\kappa^2}{4} \left( \beta \square
+ B \square^2 \right)  \right] {\ga_{\mu\rho}}^{\,,\mu\rho}
 - \,
\frac{\ka}{2} \left( \al
+ \frac{\beta}{2} + A\square + \frac{B}{2} \square \right)
\square R .
\eeq
\item[iii)] Summing up the last two equations and using
\eqref{RLin} yields
\beq
\label{R/3k}
\frac{R}{3\kappa}
\,=\,
\frac{\kappa}{12}\,T
- \frac{\kappa}{2} \left( \alpha + \frac{\beta}{3}
+ A \square + \frac{B}{3} \square \right)  \square R.
\eeq
\end{itemize}
Then, inserting \eqref{R/3k} into \eqref{deriv_hE} gives
\beq
\label{eq_mot_h(E)}
\square h_{\mu\nu}^{(E)} = - \frac{\kappa}{2} \left( T_{\mu\nu} - \frac{T}{2} \eta_{\mu\nu} \right)
\eeq
where we defined the new field
\beq
\label{def_h(E)}
h_{\mu\nu}^{(E)}
 =
 -\, \square \psi_{\mu\nu} - \frac{\beta}{B} \psi_{\mu\nu}
 + h_{\mu\nu}
  + \frac{\kappa}{2} \left( \alpha + \frac{\beta}{3}
 + A \square + \frac{B}{3} \square \right)  R\, \eta_{\mu\nu} .
\eeq

One can rewrite~\eqref{def_h(E)} in an alternative useful form
\beq
\label{solucao_1}
h_{\mu\nu}
&=&
h_{\mu\nu}^{(E)} + \left( m_{2+}^2 + m_{2-}^2
+ \square \right) \psi_{\mu\nu}
 - \frac{\kappa}{2} \left( A + \frac{B}{3} \right)
\left( \frac{3\alpha + \beta}{3A + B}
+ \square \right)  R \,\eta_{\mu\nu}\, .
\eeq

The only field which remains to be defined is the scalar $\phi$.
Equation~\eqref{R/3k} can be rewritten in the factorized form
\beq
\label{R/3k_2}
\frac{\kappa}{2} \left( A + \frac{B}{3} \right) \left( m_{0+}^2
+ \square \right) \left(  m_{0-}^2 + \square \right) R
\,=\, \frac{\kappa}{12}\,T\, ,
\eeq
where the quantities $m_{0+}^2$ and $m_{0-}^2$ satisfy
\beq
m_{0+}^2 + m_{0-}^2
&=& \frac{3\alpha + \beta}{3A + B} ,
\\
m_{0+}^2 m_{0-}^2
&=& \frac{2}{\kappa^2 (3A + B)} \,.
\label{cond2}
\eeq
It is straightforward to verify that the solution of this system
is the second relation in~\eqref{Def_masses}. Hence one can
define the scalar field
\beq
\label{def_Phi}
\phi
&=&
\frac{\kappa}{2} \left( A + \frac{B}{3} \right) R\,,
\eeq
while its equation of motion follows from \eqref{R/3k_2},
\beq
\label{eq_mot_phi}
\left( m_{0+}^2 + \square \right)
\left(  m_{0-}^2 + \square \right) \phi
\,=\, \frac{\kappa}{12} \,T\,.
\eeq

The general solution \eqref{solucao_1} of the system
\eqref{eqMotion6LinB} can be presented in the form
\beq
\label{generalsolution}
h_{\mu\nu}
&=&
h_{\mu\nu}^{(E)} + \left( m_{2+}^2 + m_{2-}^2
+ \square \right) \psi_{\mu\nu}
 - \, \eta_{\mu\nu} \left( m_{0+}^2
+ m_{0-}^2 + \square \right) \phi \,.
\eeq

Up to this point we have shown that the general solution of
\eqref{SystemAppendix} is written as a combination of three
independent fields which satisfy the equations of motion \eqref{eq_mot_psi}, \eqref{eq_mot_h(E)}
and~\eqref{eq_mot_phi}. In order to complete de proof one
has to show that the tensor fields $h_{\mu\nu}^{(E)}$ and
$\psi_{\mu\nu}$  satisfy the gauge conditions.

In terms of $\,\ga_{\mu\nu}^{(E)} = h_{\mu\nu}^{(E)}
- \frac{1}{2} h^{(E)} \eta_{\mu\nu}$, Eq.~\eqref{eq_mot_h(E)}
can be written as
\beq
\label{identities2}
\square \gamma_{\mu\nu}^{(E)}
\,=\, -\, \frac{\kappa}{2} T_{\mu\nu} \,.
\eeq
One can note that the gauge condition $\,\Ga_\mu = 0\,$ is
equivalent to ${\Om_{\mu\nu}}^{,\nu} = 0$, where
\beq
\Om_{\mu\nu}
&=&
\left[ 1 - \frac{\kappa^2}{4}
\left( \beta \square + B \square^2 \right)  \right]
\ga_{\mu\nu}
 - \, \frac{\ka}{2} \left(\al + \frac{\be}{2}
+ A\square + \frac{B}{2} \square \right) R\, \eta_{\mu\nu} \,.
\eeq
According to Eq.~\eqref{generalsolution} it follows
\beq
\label{gammaMuNu}
\ga_{\mu\nu}
&=&
\ga_{\mu\nu}^{(E)}
+ \left( \frac{\beta}{B} + \square \right) \left( \psi_{\mu\nu}
- \frac{1}{2} \eta_{\mu\nu} \psi \right)
 + \left( \frac{\si_1}{\si_2} + \square \right) \phi \eta_{\mu\nu}\,.
\eeq
By combining Eqs.~\eqref{eq_mot_psi}, \eqref{def_Phi}, \eqref{identities2}
and \eqref{gammaMuNu}, it is easy to show that
\beq
\Omega_{\mu\nu}
&=&
\ga_{\mu\nu}^{(E)}\,.
\eeq

Therefore, the gauge condition \eqref{SystemBAppendix} implies
\beq
\ga_{\mu\nu}^{(E),\nu} \,=\, 0\,.
\label{DeDonderGauge}
\eeq
Together with the equation of motion \eqref{eq_mot_h(E)},
this means $h_{\mu\nu}^{(E)}$ is the solution of linearised
general relativity in de Donder gauge.

The gauge condition for the field $\psi_{\mu\nu}$ can
be obtained by remembering that [see Eq.~\eqref{RLin}]
\beq
\ga_{\mu\nu}\,^{,\mu\nu}
&=&
\frac{1}{2}\, \square h - \frac{1}{\kappa}\,R\, .
\eeq
Taking into account \eqref{generalsolution}, \eqref{gammaMuNu}
and \eqref{DeDonderGauge} in the previous expression it can be
shown that
\beq
\left( m_{2+}^2 + m_{2-}^2 + \square\right)
\left( {\psi_{\mu\nu}}^{,\mu\nu} - \square \psi\right)  \,=\, 0 ,
\eeq
completing the proof.

The Theorem~\ref{Theorem1} can then be regarded as a corollary of the previous proposition which follows from the change of variables
\beq
&& \overline{\Psi}_{\mu\nu} \,=\,  m_{2+}^2 \psi_{\mu\nu} ,
\quad \Psi_{\mu\nu} \,=\, ( m_{2-}^2 + \square ) \psi_{\mu\nu}   ,
\\
&& \overline{\Phi} \,=\,  m_{0+}^2 \phi ,
\quad
\Phi \,=\, ( m_{0-}^2 + \square ) \phi
\eeq
in Eqs.~\eqref{Theorem_GeneralSolution}-\eqref{Theorem_Phi}.

\section*{Acknowledgements}
B.L.G. and I.Sh. are very grateful to A. Starobinsky for clarifying
discussions and criticism.
A.A. acknowledges CNPq and FAPERJ for financial support. B.L.G.
is thankful to CNPq for supporting his Ph.D. project. I.Sh. is grateful
to CNPq, FAPEMIG and ICTP for partial support of his work.
B.L.G. wishes to thank the Department of Physics at Universidade
Federal de Juiz de Fora for the kind hospitality during his visit,
when the initial version of this work was completed.


\end{document}